\documentclass[aps,eqsecnum,preprint,floats,epsf,epsfig,nofootinbib]{revtex4}
\usepackage{epsfig}

\begin{document}
\newcommand{\sla}[1]{#1\!\!\!\slash}
\def\be{\begin{eqnarray}}
\def\en{\end{eqnarray}}
\def\non{\nonumber}
\def\la{\langle}
\def\ra{\rangle}
\def\nc{N_c^{\rm eff}}
\def\vp{\varepsilon}
\def\drho{\bar\rho}
\def\deta{\bar\eta}
\def\CP{{\it CP}~}
\def\a{{\cal A}}
\def\B{{\cal B}}
\def\c{{\cal C}}
\def\d{{\cal D}}
\def\e{{\cal E}}
\def\p{{\cal P}}
\def\t{{\cal T}}
\def\up{\uparrow}
\def\dw{\downarrow}
\def\vma{{_{V-A}}}
\def\vpa{{_{V+A}}}
\def\smp{{_{S-P}}}
\def\spp{{_{S+P}}}
\def\J{{J/\psi}}
\def\ov{\overline}
\def\Lqcd{{\Lambda_{\rm QCD}}}
\def\pr{{Phys. Rev.}~}
\def\prl{{Phys. Rev. Lett.}~}
\def\pl{{Phys. Lett.}~}
\def\np{{Nucl. Phys.}~}
\def\zp{{Z. Phys.}~}
\def\lsim{ {\ \lower-1.2pt\vbox{\hbox{\rlap{$<$}\lower5pt\vbox{\hbox{$\sim$}
}}}\ } }
\def\gsim{ {\ \lower-1.2pt\vbox{\hbox{\rlap{$>$}\lower5pt\vbox{\hbox{$\sim$}
}}}\ } }

\font\el=cmbx10 scaled \magstep2{\obeylines\hfill July, 2006}

\vskip 1.5 cm

\centerline{\large\bf Production of $P$-wave Charmed Mesons in
Hadronic $B$ Decays}
\bigskip
\centerline{\bf Hai-Yang Cheng and Chun-Khiang Chua}
\medskip
\centerline{Institute of Physics, Academia Sinica}
\centerline{Taipei, Taiwan 115, Republic of China}
\medskip

\bigskip
\bigskip
\centerline{\bf Abstract}
\bigskip
{\small  Production of even-parity charmed mesons in hadronic $B$
decays is studied. Specifically,  we focus on the Cabibbo-allowed
decays $\ov B\to D^{**}\pi$ and $\bar D^{**}_sD^{(*)}$, where
$D^{**}$ denotes generically a $p$-wave charmed meson. While the
measured color-allowed decays $\ov B^0\to D^{**+}\pi^-$ are
consistent with the theoretical expectation, the experimental
observation of $B^-\to D^{**0}\pi^-$ for the broad $D^{**}$ states
is astonishing as it requires that the color-suppressed
contribution dominates over the color-allowed one, even though the
former is $1/m_b$ suppressed in the heavy quark limit. In order to
accommodate the data of $\ov B\to D^{**}\pi^-$, it is found that
the real part of  $a_2/a_1$ has a sign opposite to that in $\ov
B\to D\pi$ decays, where $a_1$ and $a_2$ are the effective
parameters for color-allowed and color-suppressed decay
amplitudes, respectively. The decay constants and form factors for
$D^{**}$ and the Isgur-Wise functions $\tau_{1/2}(\omega)$ and
$\tau_{3/2}(\omega)$ are extracted from the data of $B\to
D^{**}\pi$ decays. The Isgur-Wise functions calculated in the
covariant light-front quark model are in good agreement with
experiment. The neutral modes $\ov B^0\to D^{**0}\pi^0$ for
$D^{**}=D^*_0(2400),D'_1(2430)$ and $\ov B^0\to
D'^0_1(2430)\omega$ are predicted to have branching ratios of
order $10^{-4}$ which are also supported by isospin argument. The
decay constants of $D_{s0}^*(2317)$ and $D'_{s1}(2460)$ are
inferred from the measurements of $\ov B\to D_s^{**-}D$ to be
$58\sim 86$ MeV and $130\sim 200$ MeV, respectively. Contrary to
the decay constants $f_{D_0^*}$ and $f_{D'_1}$ which are similar
in size, the large disparity between $f_{D_{s0}^*}$ and
$f_{D'_{s1}}$ is surprising and unexpected.

\pagebreak

\section{Introduction}

The spectroscopy for the $p$-wave charmed mesons has entered a new
and exciting era since 2003. First of all, BaBar \cite{BaBarDs0}
has discovered a new narrow and light resonance $D_{s0}^*(2317)$.
The existence of a second narrow resonance $D_{sJ}(2460)$ which
can be identified with $J^P=1^+$ state was first hinted by BaBar
\cite{BaBarDs0} and then observed and established by CLEO
\cite{CLEODs1} and Belle \cite{BelleDs1}. Second, the broad
$D_0^*$ and $D_1$ resonances which are the counterpart of
$D_{s0}^*(2317)$ and $D'_{s1}(2460)$ in the non-strange charm
sector, were explored by Belle \cite{BelleD} in charged $B$ to
$D^+\pi^-\pi^-$ and $D^{*+}\pi^-\pi^-$ decays and by FOCUS
\cite{FOCUS} in photoproduction experiment.

Just like the light scalar mesons $f_0(980),a_0(980),\sigma(600)$,
the underlying structure of the above-mentioned $p$-wave charmed
mesons is not well established theoretically. Recall that before
2003, the $p$-wave states $D_{s0}^*$ and $D'_{s1}$ with $j_q=1/2$
($j_q$ being the angular momentum of the light degrees of freedom)
are predicted to be broad and decay into $DK$ and $D^*K$,
respectively, in the conventional quark model \cite{QM}. However,
the observed $D_{s0}^*(2317)$ and $D'_{s1}(2460)$ states are below
the $DK$ and $DK^*$ thresholds, respectively, and hence are very
narrow. This unexpected and surprising disparity between theory
and experiment has sparked a flurry of many theory studies. It has
been advocated that this new state is a  $DK$ molecular
\cite{Barnes} or  a $D_s\pi$ atom \cite{Szc} or a four-quark bound
state (first proposed in \cite{CH}, followed by
\cite{4quark,Bracco}).\footnote{An issue for the four-quark model
is whether there exist the conventional $c\bar s$ and $c\bar q$
states. A non-observation of a heavier and broad $0^+$ $c\bar s$
state will not support the four-quark interpretation of
$D_{s0}^{*}(2317)$. For non-strange charmed mesons, it has been
argued that $D_0^*(2308)^0$ observed by Belle \cite{BelleD} is a
four-quark state, whereas $D_0^*(2405)^{0,+}$ measured by FOCUS
\cite{FOCUS} is a normal $c\bar q$ state \cite{Bracco}; that is,
the $D_0^*$ state observed by Belle and FOCUS may not be the
same.}
On the contrary, it has been put forward that, based on heavy
quark effective theory (HQET) and chiral perturbation theory, the
newly observed $D_s(2317)$ is a $0^+$ $c\bar s$ state and that
there is a $1^+$ chiral partner with the same mass splitting with
respect to the $1^-$ state as that between the $0^+$ and $0^-$
states \cite{Bardeen,Nowak}, namely,
$m_{D'_{s1}}-m_{D_s^*}=m_{D_{s0}^*}-m_{D_s}$.

The spectra and strong, radiative decays of the $p$-wave charmed
mesons have been studied extensively (for a review, see
\cite{Colangelo,Close,Swanson}). The measurements of two-body $B$
decays into $D^{**}$ and $D_s^{**}$, where $D^{**}$ denotes a
generic even-parity charmed meson, provide very useful information
on the decay constants and form factors of the excited charmed
meson and the ratios of radiative to hadronic decay rates, for
example, $\Gamma(D_{s0}^*\to D_s^*\gamma)/\Gamma(D_{s0}^*\to
D_s\pi^0)$. Moreover, they provide an opportunity to test heavy
quark effective theory. As we shall see below, the decay
amplitudes of $B\to D^{**}\pi$ in the heavy quark limit are
governed by the Isgur-Wise functions $\tau_{1/2}(\omega)$ and
$\tau_{3/2}(\omega)$.

In the present work, we will study even-parity charmed meson
production in $B$ decays. Specifically, we focus on the
Cabibbo-allowed decays $\ov B\to D^{**}\pi$ and $\bar
D^{**}_sD^{(*)}$. \footnote{These decays have been studied
previously in
\cite{Katoch,Neubert,Castro,Kim,Lee,Chen1,Cheng03,Chen2,Jugeau,Thomas,Datta}.}
The decay $B^-\to D^{**0}\pi^-$ receives color-allowed and
color-suppressed contributions, characterized by the effective
Wilson parameters $a_1$ and $a_2$, respectively. It is important
to know whether the interference between the $a_1$ and $a_2$ terms
is constructive or destructive. In the naive factorization
approach, $a_1=c_1+c_2/N_c$ and $a_2=c_2+c_1/N_c$. In the late
80's and early 90's, the large-$N_c$ approach was very popular and
successful in explaining the hadronic weak decays of charmed
mesons \cite{Buras}.  It was widely believed by most practitioners
in the field that the $1/N_c$ expansion applies equally well to
the weak decays of $B$ mesons. Since $a_2/a_1\approx c_2/c_1\sim
-0.25$ at the renormalization scale $\mu=m_B$ in the leading
$1/N_c$ expansion, $\ov B^0\to D^+\pi^-$ is naively expected to
have a larger rate than $B^-\to D^0\pi^-$ due to the destructive
interference in the latter. However, the CLEO measurements of
$B\to D\pi$ imply the opposite \cite{CLEODpi} ! This is a very
stunning result. In order to accommodate the $B\to D\pi$ data, the
ratio $a_2/a_1$ is found to be order of $0.20\sim 0.25$. In early
2000's, the color-suppressed mode $\ov B^0\to D^0\pi^0$ is found
to be significantly larger than the theoretical expectation based
on naive factorization. For example, the measurement $\B(\ov
B^0\to D^0\pi^0)=(2.91\pm0.28)\times 10^{-4}$ \cite{PDG} is larger
than the theoretical prediction, $(0.58\sim 1.13)\times 10^{-4}$
\cite{ChengBDpi}, by a factor of $2\sim 4$. Moreover, the three
$B\to D\pi$ amplitudes form a non-flat triangle, indicating
nontrivial relative strong phases between them. As a consequence,
$a_2/a_1\approx (0.45-0.65)e^{\pm i60^\circ}$ is inferred from the
$B\to D\pi$ measurements including the neutral mode $\ov B^0\to
D^0\pi^0$ \cite{ChengBDpi,Xing,NP,LeeDpi}. The question is then
why the magnitude and phase of $a_2/a_1$ are so different from the
model expectation. To resolve this difficulty, it has been shown
in \cite{CCS,CHY} that the enhancement of the effective $a_2$ and
its strong phase can be ascribed to final-state interactions.

For $B\to D^{**}\pi$ decays, we will pay a great attention to the
relative sign of the decay constants and form factors of the
$p$-wave mesons. It turns out that in order to explain the larger
rate of $B^-\to D^{**0}\pi^-$ than $\ov B^0\to D^{**+}\pi^-$ for
$D^{**}=D_0^*$ and $D'_1$, the real part of $a_2/a_1$ for $B\to
D^{**}\pi$ has to be {\it negative} with a large magnitude.
Moreover, the color-suppressed contribution has to dominate over
the color-allowed one, in contrast to the naive expectation that
the color-suppressed amplitude is $1/m_b$ suppressed in the heavy
quark limit. This is a third surprise for the ratio $a_2/a_1$ !
The question to be addressed is why the sign of the real part of
$a_2/a_1$ flips when $B\to D\pi$ is replaced by $B\to D^{**}\pi$.

This work is organized as follows. In Sec. II, the decay constants
of $D^{**}$ and $B\to D^{**}$ form factors  within the covariant
light-front quark model are summarized. The decays $\ov B\to
D^{**}\pi$ and $\ov B\to \ov D_s^{**}D$ are studied in Secs. III
and IV, respectively. Conclusions are presented in Sec. V.

\begin{table}[h]
\caption{The masses and decay widths of even-parity charmed mesons
\cite{PDG}.  The four $p$-wave charmed meson states are denoted by
$D_0^*,D'_1,D_1$ and $D_2^*$. In the heavy quark limit, $D'_1$ has
$j=1/2$ and $D_1$ has $j=3/2$ with $j$ being the total angular
momentum of the light degrees of freedom.
 } \label{tab:mass}
\begin{ruledtabular}
\begin{tabular}{l l l  }
~~~~State~~~~ & Mass (MeV) & Width (MeV) \\
\hline
 $D^*_0(2400)^0$~\footnotemark[1] & $2352\pm50$ & $261\pm50$  \\
 $D^*_0(2400)^\pm$~\footnotemark[1] & $2403\pm14\pm35$ & $283\pm24\pm34$  \\
 $D_1(2420)^0$~\footnotemark[2] & $2421.8\pm0.8$ & $20.3^{+1.9}_{-1.8}$  \\
 $D_1(2420)^\pm$ & $2427\pm5$ & $26\pm8$  \\
 $D'_1(2430)^\pm$ & $2427\pm26\pm25$ & $384^{+107}_{-~75}\pm74$ \\
 $D_2^*(2460)^0$~\footnotemark[3] & $2462.7\pm 0.8$ & $44\pm 2$  \\
 $D_2^*(2460)^\pm$ & $2464.9\pm 3.0$ & $29\pm5$  \\
 \hline
 $D_{s0}^*(2317)$ & $2317.3\pm0.4\pm0.8$ & $<3.8$~\footnotemark[4]  \\
 $D'_{s1}(2460)$ & $2458.9\pm0.9$ & $<3.5$~\footnotemark[4]  \\
 $D_{s1}(2536)$ & $2535.35\pm0.31$ & $<2.3$  \\
 $D_{s2}^*(2573)$ & $2573.5\pm1.7$ & $15^{+5}_{-4}$  \\
\end{tabular}
\end{ruledtabular}
\footnotetext[1]{While the mass and the width of $D_0^*(2400)^0$
arise from the average of Belle \cite{BelleD} and FOCUS
\cite{FOCUS} measurements, the mass and the width of
$D_0^*(2400)^\pm$ are solely due to FOCUS \cite{FOCUS}.}
 \footnotetext[2]{Including the most recent CDF
measurements $m(D_1^0)=2421.7\pm0.7\pm0.6$ MeV and
$\Gamma(D_1^0)=20.0\pm1.7\pm1.3$ MeV \cite{CDF} to the 2005 PDG
average \cite{PDG}.}
 \footnotetext[3]{Including the most
recent CDF measurements $m(D_2^{*0})=2463.3\pm0.6\pm0.8$ MeV and
$\Gamma(D_2^{*0})=49.2\pm2.3\pm1.3$ MeV \cite{CDF} to the 2005 PDG
average \cite{PDG}.}
 \footnotetext[4]{The width limit from BaBar \cite{BaBarDs0Ds1}.}
\end{table}

\section{Decay constants and form factors}
In the quark model, the even-parity mesons are conventionally
classified according to the quantum numbers $J,L,S$: the scalar
and tensor mesons correspond to $^{2S+1}L_J=\,^3P_0$ and $^3P_2$,
respectively, and there exit two different axial-vector meson
states, namely, $^1P_1$ and $^3P_1$ which can undergo mixing if
the two constituent quarks do not have the same masses. For heavy
mesons, the heavy quark spin $S_Q$ decouples from the other
degrees of freedom in the heavy quark limit, so that $S_Q$ and the
total angular momentum of the light quark $j$ are separately good
quantum numbers.  The total angular momentum $J$ of the meson is
given by $\vec J=\vec j+\vec S_Q$ with $\vec S=\vec s+\vec S_Q$
being the total spin angular momentum. Consequently, it is more
natural to use $L^{j}_J=P_2^{3/2},P_1^{3/2},P_1^{1/2}$ and
$P_0^{1/2}$ to classify the first excited heavy meson states where
$L$ here is the orbital angular momentum of the light quark. It is
obvious that the first and last of these states are $^3P_2$ and
$^3P_0$, while \cite{IW}
 \be \label{PjJ}
 |P_1^{3/2}\ra=\sqrt{2\over 3}\,|\,^1P_1\ra+\sqrt{1\over 3}\,|\,^3P_1\ra,
 \qquad  |P_1^{1/2}\ra=-\sqrt{1\over 3}\,|\,^1P_1\ra+\sqrt{2\over
 3}\,|\,^3P_1\ra.
 \en
In the heavy quark limit, the physical eigenstates with $J^P=1^+$
are $P_1^{3/2}$ and $P_1^{1/2}$ rather than $^3P_1$ and $^1P_1$.

The masses and decay widths of even-parity (or $p$-wave) charmed
mesons $D_J^*$ and $D_{sJ}^*$ are summarized in Table
\ref{tab:mass}. It is known that in the non-charm scalar meson
sector, the quark model cannot explain why the scalar strange
meson $K_0^*(1430)$ with a mass $1412\pm 6$ MeV \cite{PDG} is
lighter than the non-strange one $a_0(1450)$ with a mass
$1474\pm19$ MeV \cite{PDG}. \footnote{Recently it has been
advocated in \cite{Maiani} that the puzzle with the relative
masses of $K_0^*(1430)$ and $a_0(1450)$ can be resolved provided
that the observed scalar nonet in the mass range 1-2 GeV is a
tetraquark nonet plus a glueball. The yet-to-be observed $q\bar q$
scalar nonet lies around 1.1 GeV.}
Likewise, it is clear from Table \ref{tab:mass} that the relation
$m(D_s^{**})>m(D^{**})$ holds except for $D_{s0}^*$ and $D_0^*$
for which we have $m(D_{s0}^*)<m(D_0^*)$. This is the place where
the conventional quark model seems not to work \footnote{There are
some attempts in understanding the mass relation
$m(D_{s0}^*)<m(D_0^*)$ by modifying the conventional potential
model. For example, one loop chiral corrections to the potential
model is considered in \cite{Taekoon}, while the potential model
in \cite{Matsuki} takes into account negative energy states of a
heavy quark in a bound state.}
and a four-quark structure for $D_0^*$ and $D_{s0}^*$ is
preferable \cite{Bracco}.

We shall use $1'^+$ and $1^+$ or $D'_1$ and $D_1$ to distinguish
between two different physical axial-vector charmed meson
states.\footnote{The notation for $D'_1(2430)$ and $D_1(2420)$ is
opposite to that in \cite{ChengBDpi}.}
The physical $1'^+$
state is primarily $P_1^{1/2}$, while $1^+$ is predominately
$P_1^{3/2}$. This is because, in the heavy quark limit, the
physical mass eigenstates $D'_1$ and $D_1$ can be identified with
$P_1^{1/2}$ and $P_1^{3/2}$, respectively. However, beyond the
heavy quark limit, there is a mixing between $P_1^{1/2}$ and
$P_1^{3/2}$, denoted by $D_1^{1/2}$ and $D_1^{3/2}$ respectively,
 \be \label{Dmixing}
 D'_1(2430) &=& D_1^{1/2}\cos\theta+D_1^{3/2}\sin\theta, \non \\
 D_1(2420) &=& -D_1^{1/2}\sin\theta+D_1^{3/2}\cos\theta.
 \en
Likewise, for strange axial-vector charmed mesons
 \be \label{Dsmixing}
 D'_{s1}(2460) &=& D_{s1}^{1/2}\cos\theta_s+D_{s1}^{3/2}\sin\theta_s, \non \\
 D_{s1}(2536) &=& -D_{s1}^{1/2}\sin\theta_s+D_{s1}^{3/2}\cos\theta_s.
 \en
Since $D_1^{1/2}$ is much broader than $D_1^{3/2}$, the decay
width of $D_1(2420)$ is sensitive to the mixing angle $\theta$.
The $D_1^{1/2}\!-\!D_1^{3/2}$ mixing angle was reported to be
 \be \label{eq:mixingangle}
 \theta=0.10\pm0.03\pm0.02\pm0.02\,{\rm rad}=(5.7\pm2.4)^\circ
 \en
by Belle through a detailed analysis of $B\to D^*\pi\pi$
\cite{BelleD}.

Since the decay $B^-\to D^{**0}\pi^-$ receives color-allowed and
color-suppressed contributions, it is important to know whether
the interference is constructive or destructive. In \cite{CCH} we
have computed the decay constants and form factors for the
ground-state $s$-wave and low-lying $p$-wave mesons within the
framework of a covariant light-front approach. \footnote{There are
many typos in the printed version of \cite{CCH} but not in the
archive version, hep-ph/0310359.}
In our approach, we first fix the vertex functions (i.e. Feynman
rules for the meson-quark-antiquark vertices) for both $s$-wave
and $p$-wave mesons. Then we are able to compute their decay
constants and form factors. Hence, the relative sign and the
factors of $i$ between two-body and three-body matrix elements can
be determined. We then adopt two different approaches to elaborate
on the heavy quark limit behavior of physical quantities: one from
top to bottom and the other from bottom to top. In the
top-to-bottom approach, we  derive the decay constants and form
factors in the covariant light-front model within HQET and obtain
model-independent heavy quark symmetry (HQS) relations. In the
bottom-to-top approach, we study the heavy quark limit behavior of
the decay constants and transition form factors of heavy mesons
and show that they do match the covariant model results based on
HQET \cite{CCH}.

\subsection{Decay constants}

The decay constants of scalar and pseudoscalar mesons are defined
by \footnote{Sometimes the decay constant of the pseudoscalar
meson is defined as $\la 0|A_\mu|P(q)\ra=f_P q_\mu$ in the
literature (see e.g. \cite{Jugeau}.) This corresponds to choosing
a non-hermitian vertex function $\Gamma=\gamma_5$ for the
pseudoscalar meson or redefining the phase of the pseudoscalar
field, namely $|P\ra\to \exp(i\phi)|P\ra$ with $\phi=\pi$. The
two-body matrix element $\la P|V_\mu|B\ra$ is still given by Eq.
(\ref{eq:ffpwave}). However, the usual soft-pion theorem
 \be
 \lim_{q\to 0}\la \pi(q)|V_\mu|B\ra=-i{\sqrt{2}\over f_\pi}\la
 0|A_\mu|B\ra,
 \en
has to be modified to
 \be
 \lim_{q\to 0}\la \pi(q)|V_\mu|B\ra={\sqrt{2}\over f_\pi}\la
 0|A_\mu|B\ra.
 \en
}
 \be \label{decaycon}
  \la 0|A_\mu|P(q)\ra &=& if_Pq_\mu, \qquad\qquad\quad \la
0|V_\mu|S(q)\ra= f_S q_\mu.
 \en
It is known that the decay constants of non-charm light scalar
mesons are smaller than that of pseudoscalar mesons as they vanish
in the SU(3) limit. The decay constants of the axial-vector
charmed mesons are defined by
 \be
 \la 0|A_\mu|D_1^{1/2}(q,\vp)\ra=\,f_{D_1^{1/2}}m_{D_1^{1/2}}\vp_\mu, \qquad
  \la 0|A_\mu|D_1^{3/2}(q,\vp)\ra=\,f_{D_1^{3/2}}m_{D_1^{3/2}}\vp_\mu.
 \en
It is known that in the heavy quark limit \cite{HQfrules},
 \be \label{fHQS}
 f_{D_1^{1/2}}=f_{D_0^*}, \qquad\quad f_{D_1^{3/2}}=0.
 \en
Since the decay constant of $D_2^*$ vanishes irrespective of heavy
quark symmetry (see below), the charmed mesons within the
multiplet $(0^+,1^+)$ or $(1'^+,2^+)$ thus have the same decay
constant.

The polarization tensor $\vp_{\mu\nu}$ of a tensor meson satisfies
the relations
 \be
 \vp_{\mu\nu}=\vp_{\nu\mu}, \qquad \vp^{\mu}_{~\mu}=0, \qquad p_\mu
 \vp^{\mu\nu}=p_\nu\vp^{\mu\nu}=0.
 \en
Therefore,
 \be
 \la 0|(V-A)_\mu|D^*_2(\vp,p)\ra=a\vp_{\mu\nu}p^\nu+b\vp^\nu_{~\nu}
 p_\mu=0.
 \en
The above relation in general follows from Lorentz covariance and
parity considerations. Hence the decay constant of the tensor
meson vanishes; that is, the tensor meson $D_2^*$ cannot be
produced from the $V-A$ current.

Using $f_D=200$ MeV, $f_{D_s}=230$ MeV and $f_{D_s^*}=230$ MeV as
input, the decay constants (in units of MeV) of $p$-wave charmed
mesons are found to be \cite{CCH}
 \be \label{decayconst}
&& f_{D_0}=86, \qquad~ f_{D_1^{1/2}}=130, \qquad f_{D_1^{3/2}}=-36, \non \\
&& f_{D_{s0}}=71, \qquad f_{D_{s1}^{1/2}}=122, \qquad
f_{D_{s1}^{3/2}}=-38,
 \en
in the covariant light-front model, where we have used the
constituent quark masses
 \be \label{eq:quarkmass}
m_{u,d}=0.26\,{\rm GeV},\qquad m_s=0.37\,{\rm GeV},\qquad
m_c=1.40\,{\rm GeV},\qquad m_b=4.64\,{\rm GeV}.
 \en
Notice that although $D_s$ has a decay constant larger than that
of $D$ as expected, it is other way around for the scalar mesons,
namely, $f_{D_0}>f_{D^*_{s0}}$. This can be seen from the
light-front quark model expression \cite{CCH}
 \be \label{eq:LFfDs}
 f_{D_s(D_{s0}^*)}\propto \int dx_2\cdots[m_c x_2\pm m_s(1-x_2)].
 \en
Since the momentum fraction $x_2$ of the strange quark in the
$D_s(D_{s0}^*)$ meson is small, its effect being constructive in
the case of $D_s$  and destructive in $D_{s0}^*$ is sizable and
explains why $f_{D_{s0}^*}/f_{D_s}\sim 0.3$ and
$f_{D^*_{s0}}<f_{D_0}$.

In principle, the decay constants of the $p$-wave strange charmed
meson $D^{**}_s$ can be extracted from the hadronic decays $B\to
\ov DD_s^{**}$ since they proceed dominantly via external
$W$-emission. In Sec. IV we shall extract the decay constants of
$D_{s0}^*$ and $D'_{s1}$ from experiment.

There are other model calculations of the $p$-wave charmed meson
decay constants. In general, these estimates are larger than ours,
(\ref{decayconst}). For example, the QCD sum rule approach in
\cite{Colangelo91} yields $f_{D_0^*}=170\pm 20\,$ MeV, while the
quark model in \cite{Veseli} predicts $f_{D_0^*}=139\pm 30\,$ MeV
and $f_{D_{s0}^*}=110\pm 18\,$ MeV. As we shall see below, some of
the decay constants for $p$-wave charmed mesons can be
phenomenologically extracted from $B\to D^{**}\pi$ and $B\to
D_s^{**}D$ decays and compared with model predictions. It turns
out that while our prediction for $f_{D_0^*}$ is smaller than
experiment, our result for $f_{D_{s0}^*}$ is in agreement with the
data.

\subsection{Form factors}
Form factors for $B\to M$ transitions with $M$ being a parity-odd
meson are given by \cite{BSW}
  \be \label{eq:ffpwave}
 \la P(p)|V_\mu|B(p_B)\ra &=& \left((p_B+p)_\mu-{m_B^2-m_{P}^2\over q^2}\,q_ \mu\right)
F_1^{BP}(q^2)+{m_B^2-m_{P}^2\over q^2}q_\mu\,F_0^{BP}(q^2), \non \\
\la V(p,\vp)|V_\mu|B(p_B)\ra &=& {2\over
m_B+m_V}\,\epsilon_{\mu\nu\alpha \beta}\vp^{*\nu}p_B^\alpha
p^\beta  V^{BV}(q^2),   \non \\
 \la V(p,\vp)|A_\mu|B(p_B)\ra &=& i\Big\{
(m_B+m_V)\vp^*_\mu A_1^{BV}(q^2)-{\vp^*\cdot p_B\over
m_B+m_V}\,(p_B+p)_\mu A_2^{BV}(q^2)    \non \\
&& -2m_V\,{\vp^*\cdot p_B\over
q^2}\,q_\mu\big[A_3^{BV}(q^2)-A_0^{BV}(q^2)\big]\Big\},
 \en
where $\epsilon_{0123}=1$, $q=p_B-p$, $F_1^{BP}(0)=F_0^{BP}(0)$,
$A_3^{BV}(0)=A_0^{BV}(0)$, and
 \be
A_3^{BV}(q^2)=\,{m_B+m_V\over 2m_V}\,A_1^{BV}(q^2)-{m_B-m_V\over
2m_V}\,A_2^{BV}(q^2).
 \en
For $B\to D^{**}$ transitions, we use
 \be
 \la D^*_0(p)|A_\mu|B(p_B)\ra &=&
-i\Bigg[\Bigg( (p_B+p)_\mu-{m_B^2-m_{D_0}^2\over q^2}\,q_
\mu\Bigg) F_1^{BD_0}(q^2) \non \\
&+& {m_B^2-m_{D_0}^2\over
q^2}q_\mu\,F_0^{BD_0}(q^2)\Bigg], \non \\
 \la D_1(p,\vp)|V_\mu|B(p_B)\ra &=&
-i\Bigg\{(m_B-m_{D_1}) \vp^*_\mu V_1^{BD_1}(q^2)  - {\vp^*\cdot
p_B\over m_B-m_{D_1}}(p_B+p)_\mu V_2^{BD_1}(q^2) \non \\
&-& 2m_{D_1} {\vp^*\cdot p_B\over
q^2}(p_B-p)_\mu\left[V_3^{BD_1}(q^2)-V_0^{BD_1}(q^2)\right]\Bigg\},
\non \\
  \la D_1(p,\vp)|A_\mu|B(p_B)\ra &=& {2\over
  m_B-m_{D_1}}\epsilon_{\mu\nu\rho\sigma}\vp^{*\nu}p_B^\rho p^\sigma
  A^{BD_1}(q^2),  \\
  \la D^*_2(p_,\vp)|V_\mu|B(p_B)\ra &=&
 h(q^2)\epsilon_{\mu\nu\rho\sigma}\vp^{*\nu\alpha}(p_B)_\alpha(p_B+p)^\rho
 (p_B-p)^\sigma,  \non \\
 \la D^*_2(p,\vp)|A_\mu|B(p_B)\ra &=& -i\Big[k(q^2)\vp^*_{\mu\nu}p_B^\nu+
 b_+(q^2)\vp^*_{\alpha\beta}p_B^\alpha p_B^\beta(p_B+p)_\mu
 \non \\
 &+& b_-(q^2)\vp^*_{\alpha\beta}p_B^\alpha
 p_B^\beta(p_B-p)_\mu\Big]. \non
 \en
with
 \be V_3^{BD_1}(q^2)=\,{m_B-m_{D_1}\over 2m_{D_1}}\,V_1^{BD_1}(q^2)-{m_B+m_{D_1}\over
2m_{D_1}}\,V_2^{BD_1}(q^2),
 \en
and $V_3^{BD_1}(0)=V_0^{BD_1}(0)$.

Note that except for the form factors $h,b_+,b_-$, all the other
form factors are dimensionless. In principle, it is better to
parametrize the form factors in such a way that they are all
positive-defined. This is the case for $B$ to $s$-wave meson
transitions, but not for all $B\to D^{**}$ transition form
factors. At any rate, the signs of various form factors can be
checked via heavy quark symmetry shown below in Eq.
(\ref{eq:formHQS}). For example, a factor of $(-i)$ is needed in
$B\to S$ transition in order for the $B\to D_0^*$ form factors
$F_{1,0}^{BD_0}$ to be positive.

Given the Feynman rules for the meson-quark-antiquark vertices
(see Table I of \cite{CCH}) in the framework of the covariant
light-front (CLF) quark model, we are able to compute the form
factors in the spacelike momentum transfer $q^2\leq 0$. Form
factors at $q^2>0$ can be obtained by first recasting them as
explicit functions of $q^2$ in the spacelike region and then
analytically continue them to the timelike region. We find that
except for the form factor $V_2^{BD^{3/2}_1}$, the momentum
dependence of form factors in the spacelike region can be well
parameterized and reproduced in the three-parameter form:
 \be \label{eq:FFpara}
 F(q^2)=\,{F(0)\over 1-a(q^2/m_{B}^2)+b(q^2/m_{B}^2)^2}\,,
 \en
for $B\to M$ transitions. The form factor $V_2^{BD^{3/2}_1}$
approaches to zero at very large $-|q^2|$ where the
three-parameter parametrization (\ref{eq:FFpara}) becomes
questionable. To overcome this difficulty, we will fit this form
factor to the form
 \be \label{eq:FFpara1}
 F(q^2)=\,{F(0)\over (1-q^2/m_{B}^2)[1-a(q^2/m_{B}^2)+b(q^2/m_{B}^2)^2]}
 \en
and achieve a substantial improvement.

Form factors for $B\to\pi$ and $B\to D^{**}$ transitions
calculated in the CLF model are listed in Tables \ref{tab:LFBtoD}.
For comparison, form factors evaluated  in the
Isgur-Scora-Grinstein-Wise (ISGW) quark model \cite{ISGW} are also
exhibited in Table \ref{tab:ISGW}. Before our work, the ISGW quark
model is the only model that can provide a systematical estimate
of the transition of a ground-state $s$-wave meson to a low-lying
$p$-wave meson. This model is based on the non-relativistic
constituent quark picture. In general, the form factors evaluated
in the original version of the ISGW model are reliable only at
$q^2=q^2_m$, the maximum momentum transfer, because the
form-factor $q^2$ dependence is proportional to
exp[$-(q^2_m-q^2)$] and hence the form factor decreases
exponentially as a function of $(q^2_m-q^2)$. This has been
improved in the ISGW2 model \cite{ISGW2} in which the form factor
has a more realistic behavior at large $(q^2_m-q^2)$ which is
expressed in terms of a certain polynomial term.

\begin{table}[t]
\caption{Form factors for $B\to \pi,D^{**}$ transitions obtained
in the covariant light-front model \cite{CCH} and fitted to the
3-parameter form Eq. (\ref{eq:FFpara}) except for the form factor
$V_2$ denoted by $^{*}$ for which the fit formula Eq.
(\ref{eq:FFpara1}) is used. All the form factors are dimensionless
except for $h,b_+,b_-$ with dimensions GeV$^{-2}$.}
\label{tab:LFBtoD}
\begin{ruledtabular}
\begin{tabular}{| c c c c c || c c c c c |}
~~~$F$~~~~~
    & $F(0)$~~~~~
    & $F(q^2_{\rm max})$~~~~
    &$a$~~~~~
    & $b$~~~~~~
& ~~~ $F$~~~~~
    & $F(0)$~~~~~
    & $F(q^2_{\rm max})$~~~~~
    & $a$~~~~~
    & $b$~~~~~~
 \\
    \hline
$F^{B\pi}_1$
    & $0.25$
    & $1.16$
    & 1.73
    & 0.95
& $F^{B\pi}_0$
    & 0.25
    & 0.86
    & 0.84
    & $0.10$
    \\
$F^{BD^*_0}_1$
    & $0.24$
    & $0.34$
    & 1.03
    & 0.27
& $F^{BD^*_0}_0$
    & 0.24
    & 0.20
    & $-0.49$
    & 0.35 \\
$A^{BD^{1/2}_1}$
    & $-0.12$
    & $-0.14$
    & 0.71
    & 0.18
&$V^{BD^{1/2}_1}_0$
    & 0.08
    & 0.13
    & 1.28
    & $-0.29$
    \\
$V^{BD^{1/2}_1}_1$
    & $-0.19$
    & $-0.13$
    & $-1.25$
    & 0.97
& $V^{BD^{1/2}_1}_2$
    & $-0.12$
    & $-0.14$
    & 0.67
    & 0.20
   \\
$A^{BD^{3/2}_1}$
    & $0.23$
    & $0.33$
    & 1.17
    & 0.39
&$V^{BD^{3/2}_1}_0$
    & $0.47$
    & $0.70$
    &  1.17
    &  0.03
    \\
$V^{BD^{3/2}_1}_1$
    & $0.55$
    & $0.51$
    & $-0.19$
    & 0.27
&$V^{BD^{3/2}_1}_2$
    & $-0.09^*$
    & $-0.17^*$
    & $2.14^*$
    & $4.21^*$
    \\

$h$
    & 0.015
    & 0.024
    & 1.67
    & 1.20
& $k$
    & 0.79
    & 1.12
    & 1.29
    & 0.93
    \\
$b_+$
    & $-0.013$
    & $-0.021$
    & 1.68
    & 0.98
& $b_-$
    & 0.011
    & 0.016
    & 1.50
    & 0.91 \\
\end{tabular}
\end{ruledtabular}
\end{table}

\begin{table}[t]
\caption{Form factors of $B\to D^{**}$ transitions calculated in
the ISGW2 model \cite{CCH}.} \label{tab:ISGW}
\begin{center}
\begin{tabular}{| c c c c c || c c c c c | }
\hline ~~~$F$~~~~~
    & $F(0)$~~~~~
    & $F(q^2_{\rm max})$~~~~
    &$a$~~~~~
    & $b$~~~~~~
    & ~~~ $F$~~~~~
    & $F(0)$~~~~~
    & $F(q^2_{\rm max})$~~~~~
    & $a$~~~~~
    & $b$~~~~~~
 \\ \hline
  $F_1^{BD_0^*}$ & 0.18 & 0.24 & 0.28 & 0.25 & $F_0^{BD_0^*}$ & 0.18 & $-0.008$ & -- & -- \\
 $A^{BD_1^{1/2}}$ & $-0.16$ & $-0.21$ & 0.87 & 0.24 &
  $V_0^{BD_1^{1/2}}$ & $0.18$ & $0.23$ & 0.89 & 0.25 \\
 $V_1^{BD_1^{1/2}}$ & $-0.19$ & $0.006$ & -- & -- & $V_2^{BD_1^{1/2}}$ &
  $-0.18$ & $-0.24$ & 0.87 & 0.24\\
 $A^{BD_1^{3/2}}$ & $0.16$ & $0.19$ & 0.46 & 0.065 &
  $V_0^{BD_1^{3/2}}$ & $0.43$ & $0.51$ & 0.54 & 0.074 \\
 $V_1^{BD_1^{3/2}}$ & $0.40$ & $0.32$ & $-0.60$ & 1.15 & $V_2^{BD_1^{3/2}}$ &
  $-0.12$ & $-0.19$ & 1.45 & 0.83 \\
 $h$ & 0.011 & 0.014 & 0.86 & 0.23 & $k$ & 0.60 & 0.68 & 0.40 & 0.68 \\
 $b_+$ & $-0.010$ & $-0.013$ & 0.86 & 0.23 & $b_-$ & 0.010 & 0.013 & 0.86 & 0.23 \\
 \hline
\end{tabular}
\end{center}
\end{table}

In the infinite quark mass limit,  all the heavy-to-heavy mesonic
decay form factors  are reduced to three universal Isgur-Wise (IW)
functions, $\xi(\omega)$ for $s$-wave to $s$-wave and
$\tau_{1/2}(\omega)$ as well as $\tau_{3/2}(\omega)$ for $s$-wave
to $p$-wave transitions, first introduced in \cite{IW}.
Specifically, the $B\to D_0^*$ and $B\to D_1^{1/2}$ form factors
are related to $\tau_{1/2}(\omega)$, while $B\to D_1^{3/2}$ and
$B\to D_2^*$ transition form factors are related to
$\tau_{3/2}(\omega)$ \cite{CCH}:
 \be \label{eq:formHQS}
 \la  D(v')|V_\mu|B(v)\ra &=& \xi(\omega)(v+v')_\mu,
 \non \\
 \la D^*(v',\vp)|V_\mu|B(v)\ra &=& -\xi(\omega)
 \epsilon_{\mu\nu\alpha\beta} \vp^{*\nu}{v'}^\alpha v^\beta, \non \\
 \la D^*(v',\vp)|A_\mu|B(v)\ra &=& i\xi(\omega)
 \Big[ (1+\omega)\vp^*_\mu-(\vp^*\cdot v)v'_\mu\Big], \non \\
 \la  D_0^*(v')|A_\mu|B(v)\ra &=& i\,2\tau_{1/2}(\omega)(v-v')_\mu,
 \non \\
 \la D_1^{1/2}(v',\vp)|V_\mu|B(v)\ra &=& -i\,2\tau_{1/2}(\omega)
 \Big[ (1-\omega)\vp^*_\mu+(\vp^*\cdot v)v'_\mu\Big],  \non \\
 \la D_1^{1/2}(v',\vp)|A_\mu|B(v)\ra &=& -2\tau_{1/2}(\omega)
 \epsilon_{\mu\nu\alpha\beta} \vp^{*\nu}{v'}^\alpha v^\beta,  \\
 \la D_1^{3/2}(v',\vp)|V_\mu|B(v)\ra &=& i{1\over\sqrt{2}}\,\tau_{3/2}(\omega)
 \Big\{ (1-\omega^2)\vp^*_\mu-(\vp^*\cdot v)[3v_\mu+(2-\omega)v'_\mu]\Big\}, \non \\
 \la D_1^{3/2}(v',\vp)|A_\mu|B(v)\ra &=& {1\over\sqrt{2}}\,\tau_{3/2}(\omega)
 (1+\omega)\epsilon_{\mu\nu\alpha\beta} \vp^{*\nu}{v'}^\alpha v^\beta,  \non \\
 \la D^*_2(v',\vp)|V_\mu|B(v)\ra &=& \sqrt{3}\,\tau_{3/2}(\omega)
 \epsilon_{\mu\nu\alpha\beta} \vp^{*\nu\gamma}v_\gamma {v'}^\alpha v^\beta, \non \\
 \la D^*_2(v',\vp)|A_\mu|B(v)\ra &=& -i\sqrt{3}\,\tau_{3/2}(\omega)
 \Big\{ (1+\omega)\vp^*_{\mu\nu}v^\nu-\vp^*_{\alpha\beta}v^\alpha v^\beta
 v'_\mu\Big\}, \non
 \en
where $\omega\equiv v\cdot v'$. For completeness, we have also
included $B\to D,D^*$ transitions in terms of the Isgur-Wise
function $\xi(\omega)$. Using the vertex functions in the heavy
quark limit given by Eqs. (4.23) and (4.24) of \cite{CCH} in
conjunction with HQET, we have derived the IW functions in the
light-front model. Their numerical expressions are given by
\cite{CCH}
 \be \label{eq:IWtheory}
 \xi(\omega) &=& 1-1.22(\omega-1)+0.85(\omega-1)^2, \non \\
 \tau_{1/2}(\omega) &=& 0.31\left(1-1.18(\omega-1)+0.87(\omega-1)^2\right), \non \\
 \tau_{3/2}(\omega) &=&
 0.61\left(1-1.73(\omega-1)+1.46(\omega-1)^2\right).
 \en
They are in good agreement with the lattice results
$\tau_{1/2}(1)=0.38\pm0.05$ and $\tau_{3/2}(1)=0.53\pm0.08$
\cite{latticeIW}. \footnote{Comparison with other model
calculations of $\tau_{1/2}(1)$ and $\tau_{3/2}(1)$ is summarized
in Table XV of \cite{CCH}. For example, based on QCD sum rules,
$\tau_{3/2}(1)=0.74\pm0.15$ is obtained \cite{Dai}.}

It is easily seen from Eq. (\ref{eq:formHQS}) that the $B\to
D^{**}$ matrix elements of weak currents vanish at the zero recoil
point $\omega=1$ owing to the orthogonality of the wave functions
of $B$ and $D^{**}$. From Eqs. (\ref{eq:ffpwave}) and
(\ref{eq:formHQS}) it is clear that the $B\to D_0^*$ and $B\to
D_1^{1/2}$ form factors in the heavy quark limit are related to
$\tau_{1/2}(\omega)$ by
 \be \label{eq:HQStau1half}
 \tau_{1/2}(\omega) &=& {\sqrt{m_Bm_{D_0^*}}\over
m_B-m_{D_0^*}}\,F_1^{BD_0^*}(q^2)=\,{\sqrt{m_Bm_{D_0^*}}\over
m_B-m_{D_0^*}}\,{F_0^{BD_0^*}(q^2)\over \left[1-{q^2\over
(m_B-m_{D_0^*})^2}\right]}
\non \\
&=& -{\sqrt{m_Bm_{D^{1/2}_1}}\over
m_B-m_{D^{1/2}_1}}\,A^{BD^{1/2}_1}(q^2)=\,
{\sqrt{m_Bm_{D^{1/2}_1}}\over m_B-m_{D^{1/2}_1}}\,V_0^{BD^{1/2}_1}(q^2)   \non \\
&=& -{\sqrt{m_Bm_{D^{1/2}_1}}\over
m_B-m_{D^{1/2}_1}}\,V_2^{BD^{1/2}_1}(q^2)=\,
-{\sqrt{m_Bm_{D^{1/2}_1}}\over
m_B-m_{D^{1/2}_1}}\,{V_1^{BD^{1/2}_1}(q^2)\over\left[1-{q^2\over
(m_B-m_{D^{1/2}_1})^2} \right]}\,.
 \en
Hence, in the heavy quark limit, we have
$V_0^{BD_1^{1/2}}(q^2)=-V_2^{BD_1^{1/2}}(q^2)=-A^{BD_1^{1/2}}(q^2)$.
Likewise, the $B\to D_1^{3/2}$ and $B\to D_2^*$ form factors are
related to $\tau_{3/2}(\omega)$ via \cite{CCH}
 \be \label{eq:HQS2}
 \tau_{3/2}(\omega) &=& -\sqrt{2\over
 m_Bm_{D^{3/2}_1}}\,{\ell_{3/2}(q^2)\over \omega^2-1}=-{1\over 3}\sqrt{2m_B^3
 \over m_{D^{3/2}_1}}\left(c^{3/2}_+(q^2)+c^{3/2}_-(q^2)\right)
 \non \\
 &=& \sqrt{2m_B^3\over m_{D^{3/2}_1}}\,{c^{3/2}_+(q^2)-c^{3/2}_-(q^2)\over
 \omega-2} = 2\,\sqrt{m_B^3m_{D^*_2}\over 3}\,h(q^2) \non \\
 &=& \sqrt{m_B\over 3m_{D^*_2}}{k(q^2)\over 1+\omega}
 =-{2\sqrt{2}\over 1+\omega}{\sqrt{m_Bm_{D^{3/2}_1}}}~q_{3/2}(q^2)
 \non \\
 &=& - \sqrt{m_B^3m_{D^*_2}\over
 3}\left(b_+(q^2)-b_-(q^2)\right),
 \en
with $\omega=(m_B^2+m_{D^{**}}^2-m_\pi^2)/(2m_Bm_{D^{**}})$,
$b_+(q^2)+b_-(q^2)=0$ and
 \be
 && \ell_{3/2}(q^2)=-(m_B-m_{D_1^{3/2}})V_1^{BD_1^{3/2}}(q^2), \qquad
 q_{3/2}(q^2)=-{A^{BD_1^{3/2}}(q^2)\over m_B-m_{D_1^{3/2}}}, \non \\
 && c_+^{3/2}(q^2)={V_2^{BD_1^{3/2}}(q^2)\over m_B-m_{D_1^{3/2}}}, \qquad\quad
 c_-^{3/2}(q^2)=-2m_{D_1^{3/2}}{V_3^{BD_1^{3/2}}(q^2)-V_0^{BD_1^{3/2}}(q^2)\over
 q^2}.
 \en
One can check that the signs of various form factors in Table
\ref{tab:LFBtoD} are in agreement with the heavy quark limit
behavior of $B\to D^{**}$ transitions, Eqs. (\ref{eq:HQStau1half})
and (\ref{eq:HQS2}).

It turns out that, among the fourteen $B\to D^{**}$ form factors,
while the covariant light-front model predictions for
$A^{BD_1^{1/2(3/2)}},V_0^{BD_1^{1/2}},V_2^{BD_1^{1/2}},h,b_+,b_-$
are in good agreement with those in the heavy quark limit, the
predictions for $F_{1,0}^{BD_0^*},V_1^{BD_1^{1/2(3/2)}}$ and $k$
at zero recoil show a large deviation from the HQS expectation.
Indeed, Eqs. (\ref{eq:HQStau1half}) and (\ref{eq:HQS2}) indicate
that except for $F_1^{BD_0^*}$, these form factors should approach
to zero when $q^2$ reaches its maximum value, a feature not borne
out in the covariant light-front calculations for finite quark
masses. This may signal that $\Lambda_{\rm QCD}/m_Q$ corrections
are particularly important in this case. Phenomenologically, it is
thus dangerous to determine all the form factors directly from the
IW functions and HQS relations since $1/m_Q$ corrections may play
an essential role for some of them and the choice of the $\beta$
parameters\footnote{$\beta$ is a wave function parameter which
governs the behavior of the phenomenological meson wave functions,
$\phi\propto {\rm exp}(-|\vec{p}^2|/2\beta^2)$. It is expected to
be of order $\Lambda_{\rm QCD}$. }
for $s$-wave and $p$-wave wave functions will affect the IW
functions.

\section{$\ov B\to D^{**}\pi$ decays}

\begin{table}[t]
\caption{Experimental branching ratio products (in units of
$10^{-4}$) of $B$ decays to $D^{**}\pi$, where
$D_0^*,D'_1,D_1,D^*_2$ stand for the charmed mesons $D_0^*(2400)$,
$D'_1(2430)$, $D_1(2420)$ and $D_2^*(2460)$, respectively. The
Cabibbo-suppressed mode $\ov B^0\to D^{*+}_2K^-$ is also
included.} \label{tab:exptBRDpi}
\begin{ruledtabular}
\begin{tabular}{l c c  c}
Mode & BaBar \cite{BaBarD,BaBarDs1K} & Belle \cite{BelleD,BelleB0Dpi}  & Average  \\
\hline
 $\B(B^-\to D^{*0}_0\pi^-)\B(D^{*0}_0\to D^+\pi^-)$ & &
 $6.1\pm0.6\pm0.9\pm1.6$ & $6.1\pm1.9$ \\
 $\B(B^-\to D'^0_1\pi^-)\B(D'^0_1\to D^{*+}\pi^-)$ & &
 $5.0\pm0.4\pm1.0\pm0.4$ & $5.0\pm1.1$ \\
 $\B(B^-\to D^0_1\pi^-)\B(D^0_1\to D^{*0}\pi^-\pi^+)$ & &
 $<0.06$ & $<0.06$ \\
 $\B(B^-\to D^0_1\pi^-)\B(D^0_1\to D^{0}\pi^-\pi^+)$ & &
 $1.85\pm0.29\pm0.35^{+0.00}_{-0.46}$ & $1.85^{+0.45}_{-0.65}$ \\
 $\B(B^-\to D^0_1\pi^-)\B(D^0_1\to D^{*+}\pi^-)$ & $5.9\pm0.3\pm1.1$ &
 $6.8\pm0.7\pm1.3\pm0.3$ & $6.2\pm0.9$ \\
 $\B(B^-\to D_2^{*0}\pi^-)\B(D^{*0}_2\to D^{*0}\pi^-\pi^+)$ & &
 $<0.22$ & $<0.22$ \\
 $\B(B^-\to D_2^{*0}\pi^-)\B(D^{*0}_2\to D^{*+}\pi^-)$ &
 $1.8\pm0.3\pm0.5$ & $1.8\pm0.3\pm0.3\pm0.2$ & $1.8\pm0.4$ \\
 $\B(B^-\to D_2^{*0}\pi^-)\B(D^{*0}_2\to D^{+}\pi^-)$ &
 $2.9\pm0.2\pm0.5$ & $3.4\pm0.3\pm0.6\pm0.4$ & $3.1\pm0.4$ \\
 \hline
 $\B(\ov B^0\to D^{*+}_0\pi^-)\B(D^{*+}_0\to D^0\pi^-)$ & &
 $0.60\pm0.13\pm0.15\pm0.22<1.20$ & $<1.20$ \\
 $\B(\ov B^0\to D'^+_1\pi^-)\B(D'^+_1\to D^{*0}\pi^+)$ & &
 $0.14\pm0.13\pm0.12^{+0.00}_{-0.10}<0.70$ & $<0.70$ \\
 $\B(\ov B^0\to D^+_1\pi^-)\B(D^+_1\to D^{*+}\pi^-\pi^+)$ & &
 $<0.33$ & $<0.33$ \\
 $\B(\ov B^0\to D^+_1\pi^-)\B(D^+_1\to D^{+}\pi^-\pi^+)$ & &
 $0.89\pm0.15\pm0.17^{+0.00}_{-0.26}$ & $0.89^{+0.23}_{-0.34}$ \\
 $\B(\ov B^0\to D^+_1\pi^-)\B(D^+_1\to D^{*0}\pi^+)$ & &
 $3.68\pm0.60^{+0.71+0.65}_{-0.40-0.30}$ & $3.7^{+1.1}_{-0.8}$ \\
 $\B(\ov B^0\to D_2^{*+}\pi^-)\B(D^{*+}_2\to D^{*+}\pi^-\pi^+)$ & &
 $<0.24$ & $<0.24$ \\
 $\B(\ov B^0\to D_2^{*+}\pi^-)\B(D^{*+}_2\to D^{*0}\pi^+)$ &
 & $2.45\pm0.42^{+0.35+0.39}_{-0.45-0.17}$ & $2.4^{+0.7}_{-0.6}$ \\
 $\B(\ov B^0\to D_2^{*+}\pi^-)\B(D^{*+}_2\to D^{0}\pi^+)$ &
 & $3.1\pm0.3\pm0.1^{+0.2}_{-0.0}$ & $3.1^{+0.4}_{-0.3}$ \\
 \hline
 $\B(\ov B^0\to D^{*+}_2K^-)\B(D_{2}^{*+}\to D^0\pi^+)$ & $0.183\pm0.051$ &
 & $0.18\pm0.05$ \\
\end{tabular}
\end{ruledtabular}
\end{table}

Given the decay constants and form factors discussed in Sec. II,
we are ready to study the $B$ decays into $p$-wave charmed mesons.
In this section we will focus on $\ov B\to D^{**}\pi$ decays. The
experimental results for the product of the branching ratios
$\B(B\to D^{**}\pi)$ and $\B(D^{**}\to$ two~particles~or
three~particles) are summarized in Table \ref{tab:exptBRDpi}.

\begin{table}[t]
\caption{Experimental branching ratios for $B\to D^{**}\pi$ decays
(in units of $10^{-4}$), where $D^{**}$ denotes a generic $p$-wave
charmed meson.
 } \label{tab:BRDpi}
\begin{center}
\begin{tabular}{|l c| c c|}
\hline
~~~~~Mode~~~~~ & ~~~~~~Expt~~~~~~~~ & ~~~~~~~~Mode~~~~~~~~ & ~~~~~~~~Expt~~~~~~~~   \\
\hline
 $B^-\to D^{*0}_0\pi^-$ &
 $9.2\pm2.9$ & $\ov B^0\to D^{*+}_0\pi^-$  & $0.90\pm0.45<1.8$ \\
 $B^-\to D'^0_1\pi^-$ &
 $7.5\pm1.7$ & $\ov B^0\to D'^+_1\pi^-$ &  $0.21^{+0.27}_{-0.30}<1.1$ \\
 $B^-\to D^0_1\pi^-$ &
 $13.5^{+1.7}_{-2.0}$ & $\ov B^0\to D_1^+\pi^-$  & $7.6^{+1.7}_{-1.4}$ \\
 $B^-\to D^{*0}_2\pi^-$ &
 $7.4\pm0.8$ & $\ov B^0\to D^{*+}_2\pi^-$  & $8.3^{+1.2}_{-1.0}$ \\
 \hline
\end{tabular}
\end{center}
\end{table}

To determine the absolute branching ratios for $B\to D^{**}\pi$,
we need some information on the branching fractions of $D^{**}$.
The decay $D_0^*$ undergoes an $s$-wave hadronic decay to $D\pi$,
while $D_1^{1/2}$ can decay into $D^*$ by $s$-wave and $d$-wave
pion emissions but only the former is allowed in the heavy quark
limit $m_c\to\infty$. Hence, we shall assume that the $D_0^*$ and
$D'_1$ widths are saturated by $D\pi$ and $D^*\pi$ respectively,
so that
 \be \label{eq:BRDD1}
 \B(D^{*0}_0\to D^+\pi^-)={2\over 3}, \qquad\quad
 \B(D'^{0}_1\to D^{*+}\pi^-)={2\over 3}.
 \en
In the heavy quark limit where the total angular momentum $j$ of
the light quark is conserved, $s$-wave $D_1^{3/2}\to D\pi$ is
prohibited by heavy quark spin symmetry. Therefore, for
$D_1(2420)$ we assume that the dominated strong decay modes are
$(D_1\to D^*\pi)_{\rm d-wave},(D\pi\pi)_{\rm
p-wave},(D^*\pi\pi)_{\rm p-wave}$. From Table \ref{tab:exptBRDpi}
it is clear that among the possible strong decays of $D_1$, the
three-body mode $D^*\pi\pi$ is suppressed relative to $D\pi\pi$.
Moreover, the analysis of $D_1(2400)\to D\pi^+\pi^-$ by Belle
\cite{BelleD1} shows that the decay mode $D_1\to D_0^*\pi$ gives
the best description. Therefore,
 \be \label{eq:BRD1}
 && \B(D_1^0\to D^0\pi^+\pi^-)\approx{2\over 3}\B(D_1^0\to D_0^{*+}\pi^-),
 \qquad \B(D_1^0\to D^{*+}\pi^-,D_0^{*+}\pi^-) = {2\over 3}.
 \en
The tensor meson $D_2^*$ decays into $D^*$ or $D$ via $d$-wave
pion emission. Since the production of $D^*\pi\pi$ in $D_2^*$
decay is very suppressed, we take
 \be \label{eq:BRD2}
\B(D_2^{*0}\to D^{*+}\pi^-,D^+\pi^-) = {2\over 3}.
 \en
In heavy quark effective theory, it is expected that
 \be
 {\Gamma(D_2^{*0}\to D^+\pi^-)\over \Gamma(D_2^{*0}\to
 D^{*+}\pi^-)}={2\over 3}\,{m_D\over m_{D^*}}\,\left({p_c(D_2^*\to
 D\pi)\over p_c(D_2^*\to D^*\pi)}\right)^5=2.3\,,
 \en
in excellent agreement with the direct measured value of $2.3\pm
0.6$ \cite{PDG}. Applying Eqs. (\ref{eq:BRDD1})-(\ref{eq:BRD2}),
the absolute branching ratios of $B\to D^{**}\pi$ are shown in
Table \ref{tab:BRDpi}. Note that in the factorization approach it
is expected that
 \be
 {\B(\ov B^0\to D_2^{*+}K^-) \over \B(\ov B^0\to D_2^{*+}\pi^-)}=
 \sin\theta_C^2,
 \en
with $\theta_C$ being the Cabibbo angle. From Table
\ref{tab:exptBRDpi} we see that this relation is well satisfied
experimentally.

We shall study $B\to D^{**}\pi$ decays within the framework of
generalized factorization in which the hadronic decay amplitude is
expressed in terms of factorizable contributions multiplied by the
{\it universal} (i.e. process independent) effective parameters
$a_i$ that are renormalization scale and scheme independent. Apart
from a common factor of $G_FV_{cb}V_{ud}^*/\sqrt{2}$, the
factorizable amplitudes for $B^-\to D^{**0}\pi^-$ read
 \be \label{eq:BtoDpi}
 A(B^-\to D_0^{*0}\pi^-) &=& -a_1
 f_\pi(m_B^2-m_{D_0}^2)F_0^{BD_0}(m_\pi^2)+
 a_2 f_{D_0}(m_B^2-m_\pi^2)F_0^{B\pi}(m_{D_0}^2), \non \\
 A(B^-\to D'^0_1\pi^-) &=&
  2(\vp^*\cdot p_B)\Bigg\{a_1f_\pi V_0^{BD'_1}(m_\pi^2)m_{D'_1} -
  a_2 f_{D'_1}F_1^{B\pi}(m_{D'_1}^2)m_{D'_1}\Bigg\}, \non \\
  A(B^-\to D_1^0\pi^-) &=&
  2(\vp^*\cdot p_B)\Bigg\{a_1f_\pi V_0^{BD_1}(m_\pi^2)m_{D_1} -
  a_2 f_{D_1}F_1^{B\pi}(m_{D_1}^2)m_{D_1}\Bigg\}, \non \\
 A(B^-\to D^{*0}_2\pi^-) &=& -a_1f_\pi\,\vp^*_{\mu\nu}p_B^\mu p_B^\nu\,\left[
 k(m_\pi^2)+b_+(m_\pi^2)(m_B^2-m_{D_2}^2)+b_-(m_\pi^2)m_\pi^2\right],
 \en
with
 \be
 V_0^{BD'_1}m_{D'_1} &=& V_0^{BD_1^{1/2}}m_{D_1^{1/2}}\cos\theta+
 V_0^{BD_1^{3/2}}m_{D_1^{3/2}}\sin\theta, \non \\
 V_0^{BD_1}m_{D_1} &=& -V_0^{BD_1^{1/2}}m_{D_1^{1/2}}\sin\theta+
 V_0^{BD_1^{3/2}}m_{D_1^{3/2}}\cos\theta, \non \\
 f_{D'_1}m_{D'_1} &=& f_{D_1^{1/2}}m_{D_1^{1/2}}\cos\theta+
 f_{D_1^{3/2}}m_{D_1^{3/2}}\sin\theta, \non \\
 f_{D_1}m_{D_1} &=& -f_{D_1^{1/2}}m_{D_1^{1/2}}\sin\theta+
 f_{D_1^{3/2}}m_{D_1^{3/2}}\cos\theta.
 \en
The decay amplitudes for $\ov B^0\to D^{**+}\pi^-$  can be
obtained from $A(B^-\to D^{**0}\pi^-)$ by setting $a_2=0$.
\footnote{It is customary to neglect the $W$-exchange
contributions to $\ov B^0\to D^{**+}\pi^-$ and $\ov B^0\to
D^{**0}\pi^0$.}
Note that except $B^-\to D_2^{*0}\pi^-$ all other decay modes
receive contributions from color-suppressed internal $W$-emission.
In the heavy quark limit,
 \be \label{eq:BtoDpiHQ}
 A(B^-\to D_0^{*0}\pi^-) &=& -2a_1
 f_\pi\sqrt{m_Bm_{D_0}}(m_B+m_{D_0})(\omega-1)\tau_{1/2}(\omega)  \non \\
 && + 2a_2 f_{D_0}F_0^{B\pi}(m_{D_0}^2)(m_B^2-m_\pi^2), \non \\
 A(B^-\to D'^0_1\pi^-) &=&  2(\vp^*\cdot v)\Bigg\{a_1f_\pi
 \sqrt{m_Bm_{D'_1}}(m_B-m_{D_1^{1/2}})\tau_{1/2}(\omega) \non \\
 && - a_2\,f_{D_1^{1/2}}F_1^{B\pi}(m_{D_1^{1/2}}^2)m_Bm_{D_1^{1/2}}\Bigg\}, \non \\
  A(B^-\to D_1^0\pi^-) &=& \sqrt{2}a_1(\vp^*\cdot v)
  f_\pi\sqrt{m_Bm_{D_1}}(m_B-m_{D_1^{3/2}})(\omega+1)\tau_{3/2}(\omega),  \non \\
 A(B^-\to D^{*0}_2\pi^-) &=& -\sqrt{3}a_1f_\pi\,\vp^*_{\mu\nu}v^\mu
 v^\nu\sqrt{m_Bm_{D_2}}(m_B+m_{D_2})\tau_{3/2}(\omega),
 \en
where $\omega=(m_B^2+m_{D^{**}}^2-m_\pi^2)/(2m_Bm_{D^{**}})$.

The decay rates are given by\footnote{ Because the scalar
resonances $D_0^*$ and $D'_1$ have widths of order 300 MeV, we
have checked the finite width effects on their production in $B$
decays and found that the conventional narrow width approximation
is accurate enough to describe the production of broad resonances
owing to the large energy released in hadronic two-body decays of
$B$ mesons \cite{Cheng03}.}
 \be \label{eq:decayrate}
 \Gamma(B\to D_0^*\pi) &=& {p_c\over 8\pi m_B^2}|A(B\to
 D_0^*\pi)|^2, \non \\
 \Gamma(B\to D_1^{(')}\pi) &=& {p^3_c\over 8\pi m_{D_1}^2}|A(B\to
 D_1^{(')}\pi)/(\vp^*\cdot p_B)|^2, \non \\
 \Gamma(B\to D_2^*\pi) &=& {p_c^5\over 12\pi m_{D_2}^2}\left({m_B\over
 m_{D_2}}\right)^2|M(B\to D_2^*\pi)|^2,
 \en
where $A(B\to D^*_2\pi)= \vp^*_{\mu\nu}p_B^\mu p_B^\nu\,M(B\to
D^*_2\pi)$ and $p_c$ is the c.m. momentum of the pion. From Eqs.
(\ref{eq:BtoDpiHQ}) and (\ref{eq:decayrate}), we obtain
 \be
 && \Gamma(\ov B^0\to D_0^{*+}\pi^-)=\Gamma(\ov B^0\to D'^+_1\pi^-)=
 {G_F\over
 16\pi}|V_{cb}V^*_{ud}|^2a_1^2f_\pi^2m_B^3\,{(1-r)^5(1+r)^3\over
 2r}|\tau_{1/2}(\omega)|^2, \non \\
  && \Gamma(\ov B^0\to D^+_1\pi^-)=\Gamma(\ov B^0\to D^{*+}_2\pi^-)=
 {G_F\over
 16\pi}|V_{cb}V^*_{ud}|^2a_1^2f_\pi^2m_B^3\,{(1-r)^5(1+r)^7\over
 16r^3}|\tau_{3/2}(\omega)|^2, \non \\
  && \Gamma(B^-\to D^0_1\pi^-)=\Gamma(\ov B^0\to D^+_1\pi^-),
  \qquad\quad \Gamma(B^-\to D^{*0}_2\pi^-)=\Gamma(\ov B^0\to
  D^{*+}_2\pi^-),
 \en
in the heavy quark limit, where $r=m_{D^{**}}/m_B$. It is evident
from Table \ref{tab:exptBRDpi} that the HQS relations $\Gamma(\ov
B^0\to D^{*+}_2\pi^-)=\Gamma(B^-\to D^{*0}_2\pi^-)$ and
$\Gamma(\ov B^0\to D^+_1\pi^-)=\Gamma(\ov B^0\to D^{*+}_2\pi^-)$
are respected by experiment, while the relation $\Gamma(\ov B^0\to
D^+_1\pi^-)=\Gamma(B^-\to D^0_1\pi^-)$ is not satisfied, implying
the importance of color-suppressed contributions which vanish in
the heavy quark limit.

From Eq. (\ref{eq:BtoDpi}) we see that apart from the coefficients
$a_1$ and $a_2$, the color-allowed and color-suppressed amplitudes
for $B^-\to \{D_0^{*0},D'^0_1,D^0_1\}\pi^-$ have opposite signs,
\footnote{ We disagree with \cite{Jugeau} on the signs. If we
follow \cite{Jugeau} to define $\la P(q)|A_\mu|0\ra=f_P q_\mu$ and
$\la D_1^{1/2}(p,\vp)|A_\mu|0\ra=-f_{D_{1/2}}m_{D_{1/2}}\vp_\mu$
for decay constants, we find that the matrix elements $\la
D_0^*|A_\mu|B\ra$ and $\la D_1^{1/2}|V_\mu|B\ra$ should have signs
opposite to that given in Eq. (28) of \cite{Jugeau}.}
in contrast to the case of $B\to D\pi$ decays.

\subsection{Color-allowed $\ov B\to D^{**}\pi$ decays}

We first discuss the color-allowed $\ov B^0\to D^{**+}\pi^-$
decays governed by the parameter $a_1$. This is the place where
the calculations are considered to be more robust. To proceed, we
first fix $a_1$ to be 0.88. In Table \ref{tab:theoryBRDpi} we show
the predictions of $\B(\ov B\to D^{**+}\pi^-)$ in the covariant
light-front model for $B\to D^{**}$ form factors and its extension
to the heavy quark limit with the IW functions given by Eq.
(\ref{eq:IWtheory}). \footnote{The predicted rates for
$D_0^{*+}\pi^-$ and $D_2^{*+}\pi^-$ are somewhat different in the
covariant model and its heavy quark limit extension. This is
because the form factors $F^{BD_0^*}(q^2)$ and $k(q^2)$ do not
respect the HQS relations (\ref{eq:HQStau1half}) and
(\ref{eq:HQS2}) satisfactorily.}
It is evident that the color-allowed modes are ``normal". To
illustrate this point, we consider the decay amplitudes in the
heavy quark limit given by Eq. (\ref{eq:BtoDpiHQ}). We see that
the color-allowed $a_1$ amplitudes for $D_0^*$ and $D'_1$
production are suppressed relative to that for $D_1$ and $D_2^*$
production because of the smallness of
$\tau_{1/2}(\omega)/\tau_{3/2}(\omega)$, $(\omega-1)/(\omega+1)$
and $(m_B-m_{D_1})/(m_B+m_{D_1})$. Note that the first three modes
in Table \ref{tab:theoryBRDpi} prefer an $a_1$ smaller than unity,
whereas the $D_2^{*+}\pi^-$ channel favors an $a_1$ close to
unity.

\begin{table}[t]
\caption{The predicted branching ratios for $\ov B^0\to
D^{**+}\pi^-$ decays (in units of $10^{-4}$) calculated in the
covariant light-front model and its extension to the heavy quark
limit (denoted by HQS). The parameter $a_1$ is taken to be
$a_1=0.88$. Experimental results are taken from Table
\ref{tab:BRDpi}.
 } \label{tab:theoryBRDpi}
\begin{center}
\begin{tabular}{|l c c c|}
\hline
Mode & ~~~~Theory~~~~~~ & ~~~~~~HQS~~~~~~ & ~~~~~~Expt~~~~~~~~ \\
\hline
 $\ov B^0\to D^{*+}_0\pi^-$  & 3.1 & 1.7 & $<1.8$ \\
 $\ov B^0\to D'^+_1\pi^-$ &  0.8 & 1.5 & $<1.1$ \\
 $\ov B^0\to D_1^+\pi^-$  & 10.4 & 11.1 & $7.6^{+1.7}_{-1.4}$ \\
 $\ov B^0\to D^{*+}_2\pi^-$  & 6.9 & 10.8 & $8.3^{+1.2}_{-1.0}$ \\
 \hline
\end{tabular}
\end{center}
\end{table}

If we treat the $B\to D_0^*,D'_1,D_1$ transition form factors as
unknown parameters, we can determine them from the data. In order
to satisfy the constraint $\B(\ov B^0\to D_0^{*+}\pi^-)<1.8\times
10^{-4}$, it follows that the $B\to D_0^*$ form factor is
constrained to be
 \be \label{eq:ffBD0}
 F_0^{BD_0^*}(0)\lsim 0.18\,.
 \en
This is smaller than the CLF prediction, $F^{BD_0^*}(0)=0.24$ (cf
Table \ref{tab:LFBtoD}). From the measurements of $\ov B^0\to
D'^+_1\pi^-,D^+_1\pi^-$ we obtain
 \be \label{eq:ffBD1}
 V_0^{BD'_1}(0)\lsim 0.15, \qquad\quad
 V_0^{BD_1}(0)=0.39^{+0.05}_{-0.03}.
 \en
Taking $V_0^{BD'_1}(0)=0.15$ and using the experimental central
value for the $D'_1-D_1$ mixing angle $\theta=5.7^\circ$ [Eq.
(\ref{eq:mixingangle})], we find
 \be \label{eq:ffBD1a}
 && V_0^{BD_1^{1/2}}(0)=0.11^{+0.00}_{-0.04}, \qquad \quad
 V_0^{BD_1^{3/2}}(0)=0.41\pm0.04,
 \en
to be compared with the CLF model predictions (see Table
\ref{tab:LFBtoD}): $V_0^{BD_1^{1/2}}(0)=0.08$ and
$V_0^{BD_1^{3/2}}(0)=0.47$.

The IW functions $\tau_{1/2}(\omega)$ and $\tau_{3/2}(\omega)$ can
be extracted from the data of $\ov B^0\to D^{**+}\pi^-$:
 \be \label{eq:ExptIW}
\ov B^0\to D^{*+}_0\pi^- \Rightarrow \qquad\quad |\tau_{1/2}(1.36)| &<& 0.22\,,  \non \\
\ov B^0\to D'^{+}_1\pi^- \Rightarrow \qquad\quad |\tau_{1/2}(1.32)| &<& 0.19\,,   \\
\ov B^0\to D^{+}_1\pi^- \Rightarrow \qquad\quad |\tau_{3/2}(1.32)| &=& 0.30\pm0.03\,,  \non \\
\ov B^0\to D^{*+}_2\pi^- \Rightarrow \qquad\quad
|\tau_{3/2}(1.31)| &=& 0.33^{+0.02}_{-0.03}\,. \non
 \en
As stressed in passing, the HQS relation $\Gamma(\ov B^0\to
D^+_1\pi^-)=\Gamma(B^-\to D^0_1\pi^-)$ is badly broken and hence
$\tau_{3/2}(\omega)$ cannot be reliably extracted from
$D_1^+\pi^-$ production. Our predictions \cite{CCH}
 \be
 && \tau_{1/2}(1)=0.31, \qquad \tau_{1/2}(1.32)=0.22, \qquad \tau_{1/2}(1.36)=0.21,
 \non \\
 && \tau_{3/2}(1)=0.61,\qquad \tau_{3/2}(1.31)=0.37
 \en
are in good agreement with the phenomenological determination
(\ref{eq:ExptIW}) and the lattice calculations \cite{latticeIW}.
For comparison, the phenomenological determination of IW functions
in \cite{Jugeau} is given by
 \be
 \tau_{1/2}(1)<0.26, &\qquad& \tau_{1/2}(1.32)<0.20, \non \\
 \tau_{3/2}(1)=0.46\pm0.18, &\qquad&
 \tau_{3/2}(1.31)=0.31\pm0.12\,.
 \en

In short, Eqs. (\ref{eq:ffBD0}),
(\ref{eq:ffBD0})-(\ref{eq:ffBD1a}) and (\ref{eq:ExptIW}) are the
main results in this subsection.

\subsection{Class-III $B^-\to D^{**0}\pi^-$ decays}

We next turn to the so-called class-III decays $B^-\to
D^{**0}\pi^-$ that receive both color-allowed and color-suppressed
contributions. The experimental observation that the production of
broad $D^{**}$ states in charged $B$ decays is more than a factor
of five larger than that produced in neutral $B$ decays (Table
\ref{tab:BRDpi}) is astonishing as it is naively expected to be a
factor of two difference at most. For example, $D_0^{*0}\pi^-$ and
$D_0^{*+}\pi^-$ rates are predicted to be similar in
\cite{Chen1,Katoch}, while $D'^0_1\pi^-$ is predicted to be even
smaller than $D'^+_1\pi^-$ in \cite{Cheng03}. The constructive
interference in $B^-\to \{D_0^{*0},D'^0_1\}\pi^-$ decays requests
that the relative sign between the real parts of $a_1$ and $a_2$
be {\it negative} as noticed in passing. Moreover, if we take the
Belle measurements of $\ov B\to \{D_0^{*},D'_1\}\pi^-$ seriously,
it will imply a color-suppressed contribution larger than the
color-allowed one, even though the former is $1/m_B$ suppressed in
the heavy quark limit ! Since the color-allowed modes have been
shown to be ``normal" before, anything unusual must arise from the
color-suppressed contributions.

Before proceeding, we need to specify the $a_2$ parameter. Recall
that $|a_1|=0.88\pm0.06$, $|a_2|=0.47\pm0.06$ and
$a_2/a_1=(0.53\pm0.06)\exp(i59^\circ)$ are obtained in
\cite{ChengBDpi} by a fit to the data of $B\to D\pi$. Owing to the
missing $\ov B^0\to D^{**0}\pi^0$ decays, one can only determine
$|1-xa_2/a_1|$ from the measurements of $B^-\to D^{**0}\pi^-$ and
$\ov B^0\to D^{**+}\pi^-$, where
$x(D_0^*\pi)=f_{D_0}(m_B^2-m_\pi^2)F_0^{B\pi}(m_{D_0)}^2)/
[f_\pi(m_B^2-m_{D_0}^2)F_0^{BD_0}(m_\pi^2)]$, for example. That
is, a determination of the relative strong phase between $a_1$ and
$a_2$ has to await the measurement of the neutral mode $\ov B^0\to
D^{**+}\pi^-$. For the present purpose, we  shall choose
$a_1=0.88$ and $a_2=-0.47$ without considering their relative
strong phase. Using this set of effective Wilson coefficients, we
obtain
 \be
 \B(B^-\to D_2^{*0}\pi^-)=7.6\times 10^{-4}, \qquad
 \B(\ov B^0\to D_2^{*+}\pi^-)=6.9\times 10^{-4},
 \en
for $\ov B\to D_2^*\pi^-$ decays, which are in agreement with
experiment.

In order to accommodate the experimental observation $\B(B^-\to
D_0^{*0}\pi^-)\gsim 5\B(\ov B^0\to D_0^{*+}\pi^-)$, the decay
constant of $D_0^*$ cannot be too small. A fit to the $B^-\to
D^{*0}_0\pi^-$ rate yields
 \be \label{eq:fD0}
 f_{D_0^*}=148^{+40}_{-46}\,{\rm MeV},
 \en
where $F_0^{BD_0}(0)$ has been set to 0.18 [see Eq.
(\ref{eq:fD0})]. This value of the $D_0^*$ decay constant is
larger than our CFL prediction $f_{D_0^*}=86$ MeV [cf Eq.
(\ref{decayconst})]. Putting the form factors (\ref{eq:ffBD1})
back to Eq. (\ref{eq:BtoDpi}) (with $V_0^{BD'_1}(0)=0.15$) and
fitting to the measured rates of $B^-\to D'^0_1\pi^-,D^0_1\pi^-$
give rise to
 \be \label{eq:fD1}
 f_{D'_1}=151^{+27}_{-30}\,{\rm MeV}, \qquad\quad
 f_{D_1}=73^{+21}_{-22}\,{\rm MeV}.
 \en
Applying the experimental central value for the $D'_1-D_1$ mixing
angle $\theta=5.7^\circ$, we find
 \be \label{eq:ff&fD1}
 && f_{D_1^{1/2}}=143^{+29}_{-32}\,{\rm MeV}, \qquad\quad~~
 f_{D_1^{3/2}}=88^{+24}_{-25}\,{\rm MeV}.
 \en
Eqs. (\ref{eq:fD0})-(\ref{eq:ff&fD1}) are the main results in this
subsection.

It should be stressed that if $|a_2|$ is chosen to be smaller, say
$a_2=-0.30$, then the decay constants $f_{D_0}$, $f_{D'_1}$ and
$f_{D_1}$ all have to be scaled up by a factor of 0.47/0.30. This
will lead to a decay constant of $D_0^*$ larger than that of the
$D$ meson. This is in contradiction to the observtion that
$f_{D_0^*}$ should be smaller than $f_D$ (see Eq.
(\ref{eq:LFfDs})). Hence, $|a_2/a_1|$ is preferred to be larger
than naive expectation.

Now we can get into more detail about the relative strength of
color-allowed and color-suppressed amplitudes in $B^-\to
\{D_0^{*0},D_1^0\}\pi^-$ decays. Since $f_{D_0},f_{D'_1}\gsim
f_\pi$, $F_0^{B\pi}(m_{D_0}^2)=0.30\gg F_0^{BD_0}(m_\pi^2)\lsim
0.18$ and $F_1^{BD'_1}(m_{D'^2_1})=0.37\gg
V_0^{BD'_1}(m_\pi^2)\lsim 0.15$ it is evident that the
color-suppressed amplitude is slightly larger than the
color-allowed one for $|a_2/a_1|=0.53$. In contrast, internal
$W$-emission is suppressed for $D_1$ and $D_2^*$ productions.
Under the factorization approximation, the color-suppression
amplitude is prohibited in $B\to D_2^*\pi$. It also vanishes in
$B\to D_1\pi$ in the heavy quark limit as $\theta\to 0$ and
$f_{D^{3/2}_1}\to 0$. Therefore, it is expected that $\Gamma(\ov
B^0\to D^{*+}_2\pi^-)=\Gamma(B^-\to D^{*0}_2\pi^-)$ in general and
$\Gamma(\ov B^0\to D^{+}_1\pi^-)=\Gamma(B^-\to D^{0}_1\pi^-)$ in
the heavy quark limit.

Comparing Eq. (\ref{eq:ff&fD1}) with Table \ref{tab:LFBtoD} and
Eq. (\ref{decayconst}), it is clear that form factors and decay
constants extracted from the data are consistent with the CLF
model calculations except for the decay constant $f_{D_1^{3/2}}$
which needs to be positive. This can be seen from Eq.
(\ref{eq:BtoDpi}) that in order to account for $\B(B^-\to
D_1^0\pi^-)>\B(\ov B^0\to D^+_1\pi^-)$, it is necessary to have a
constructive interference between color-allowed and
color-suppressed $W$-emission amplitudes. This in turn implies a
positive decay constant for $D_1^{3/2}$, i.e. $f_{D_1^{3/2}}>0$
(This is most obvious in the heavy quark limit where $\theta\to
0$.) It is not clear to us why our light-front model prediction
with a negative $f_{D_1^{3/2}}$ does not work. Of course, it is
possible that $a_2$ is process dependent and for some reason
$a_2/a_1$ becomes positive for $B^-\to D_1^0\pi^-$. Then
$f_{D_1^{3/2}}$ will be negative.

It is worth mentioning that the ratio
 \be
 R={\B(B^-\to D_2^*(2460)^0\pi^-)\over \B(B^-\to D_1(2420)^0\pi^-)}
 \en
is measured to be $0.54\pm0.18$ by Belle \cite{BelleD1}.
\footnote{The early Belle result $R=0.77\pm0.15$ \cite{BelleD}
does not take into account the contribution from $D_1(2400)\to
D\pi^+\pi^-$, so are the results $0.80\pm0.07\pm0.16$ obtained by
BaBar \cite{BaBarD} and $1.8\pm0.8$ by CLEO~\cite{Galik}.}
In soft-collinear effective theory, the equality of branching
ratios and strong phases
 \be
 {\B(\ov B\to D_2^*M)\over \B(\ov B\to D_1M)}=1, \qquad\quad
 \phi^{D_2^*M}=\phi^{D_1M}
 \en
holds in the heavy quark limit for both color-allowed and color
suppressed modes, where $M$ is a light meson \cite{Mantry}. The
early prediction by Neubert \cite{Neubert} yields a value of 0.35.

\subsection{Color-suppressed $\ov B^0\to D^{**0}\pi^0$ and $\ov B^0\to D'^0_1\omega$ decays}

The factorizable $\ov B^0\to D^{**0}\pi^0$ amplitudes are given by
 \be \label{eq:BtoD0pi0}
 A(\ov B^0\to D_0^{*0}\pi^0) &=& -{a_2\over\sqrt{2}}
 f_{D_0}(m_B^2-m_\pi^2)F_0^{B\pi}(m_{D_0}^2), \non \\
 A(\ov B^0\to D'^0_1\pi^0) &=&
  \sqrt{2}a_2 f_{D'_1}F_1^{B\pi}(m_{D'_1}^2)m_{D'_1}(\vp^*\cdot p_B), \non \\
 A(\ov B^0\to D_1^0\pi^0) &=&
  \sqrt{2}a_2f_{D_1}F_1^{B\pi}(m_{D_1}^2)m_{D_1}(\vp^*\cdot p_B).
 \en
They satisfy the isospin triangle relation
 \be \label{eq:isoBDpi}
 A(\ov B^0\to D^{**+}\pi^-)=\sqrt{2}A(\ov B^0\to D^{**0}\pi^0)+A(B^-\to D^{**0}\pi^-).
 \en
Assuming no relative strong phases between $D^{**+}\pi^-$,
$D^{**0}\pi$ and $D^{**0}\pi^0$ amplitudes (i.e. these three
amplitudes are parallel or antiparallel to each other) and using
the experimental data from Table \ref{tab:exptBRDpi}, then the
above isospin relation leads to
 \be \label{eq:isospinBR}
  && \B(\ov B^0\to D^{*0}_0\pi^0)=(1.9\pm1.0)\times 10^{-4}, \qquad
 \B(\ov B^0\to D'^{0}_1\pi^0)=(2.4\pm0.9)\times 10^{-4}, \non \\
 && \B(\ov B^0\to D^{0}_1\pi^0)=(3.0\pm2.9)\times 10^{-5}.
 \en
Since in reality there should be some relative strong phases
between the aforementioned three amplitudes, the above predictions
for color-suppressed modes can be considered as the lower bounds
and should be robust.  At the 90\% confidence level, we have
 \be
 && \B(\ov B^0\to D^{*0}_0\pi^0)>0.6\times 10^{-4}, \qquad
 \B(\ov B^0\to D'^{0}_1\pi^0)>1.1\times 10^{-4}.
 \en

Using $a_2=-0.47$ and the decay constants given in (\ref{eq:fD0})
and (\ref{eq:fD1}), a direct calculation of the branching ratios
of the neutral modes yields
 \be
 && \B(\ov B^0\to D^{*0}_0\pi^0)=(1.4^{+0.9}_{-0.7})\times 10^{-4}, \qquad
 \B(\ov B^0\to D'^{0}_1\pi^0)=(1.4\pm0.5)\times 10^{-4}, \non \\
 && \B(\ov B^0\to D^{0}_1\pi^0)=(3.2^{+2.1}_{-1.6})\times 10^{-5},
 \en
which are similar to the model-independent results
(\ref{eq:isospinBR}) derived from isospin argument. It is
important to measure these modes to see if the production of the
broad $D^{**0}$ states in neutral $B$ decays is not color and
$1/m_b$ suppressed. Also the isospin relation Eq.
(\ref{eq:isoBDpi}) will enable us to determine the relative strong
phases in $\ov B\to D^{**}\pi$ decays.

Two remarks are in order:  (i) Contrary to $\ov B\to D\pi$ decays
where $D^0\pi^0$ rates are suppressed by one order of magnitude
compared to $D^+\pi^-$, the $D^{**0}\pi^0$ rates here are
comparable to $D^{**+}\pi^-$ for broad $D^{**}$ states as the
color-suppressed amplitude is larger than the color-allowed one.
(ii) Although the $\ov B^0$ decay into $D_2^{*0}\pi^0$ is
prohibited under the factorization hypothesis, nevertheless it can
be induced via final-state interactions (FSIs) and/or
nonfactorizable contributions. In soft-collinear effective theory,
this decay receives a factorizable contribution at the leading
nonvanishing order in $\Lambda/m_B$ \cite{Mantry}.

Although the class-III decays $\ov B^0\to D^{**0}\pi^0$ have not
been observed so far, there exists one neutral mode $\ov B^0\to
D'^0_1\omega$ that can be inferred from a recent measurement of
$\ov B^0\to D^{*+}\omega\pi^-$ decays by BaBar
\cite{BaBarD1omega}. There is an enhancement for $D^*\pi$ masses
broadly distributed around 2.5 GeV. Assuming that the enhancement
is actually due to $\ov B^0\to D_1'\omega$ followed by $D'_1\to
D^{*+}\pi^-$, BaBar obtained \cite{BaBarD1omega}
 \be
 \B(\ov B^0\to D'^0_1\omega)\B(D'_1\to
 D^{*+}\pi^-)=(4.1\pm1.2\pm0.4\pm1.0)\times 10^{-4}.
 \en
It follows from Eq. (\ref{eq:BRDD1}) that
 \be
 \B(\ov B^0\to D'^0_1\omega)=(6.2\pm2.4)\times 10^{-4}.
 \en
Theoretically, the decay amplitude of $\ov B^0\to D'^0_1\omega$ is
given by
 \be
 A[\ov B^0\to D'^0_1(\vp_{D'_1},p_{D'_{1}})\omega(\vp_{\omega},p_{\omega})]
 &=& {1\over\sqrt{2}}
\vp_{D'_1}^{*\mu}\vp_{\omega}^{*\nu}[S_1 g_{\mu\nu}+S_2(p_B)_\mu
(p_B)_\nu+iS_3\epsilon_{\mu\nu\alpha\beta}p_{D'_{1}}^\alpha
p_{\omega}^\beta], \non
 \en
with (apart from a common factor of $G_FV_{cb}V_{ud}^*/\sqrt{2}$)
 \be
 S_1 &=& a_2(m_B+m_{\omega})m_{D'_{1}}f_{D'_{1}}
 A_1^{B\omega}(m_{D'_{1}}^2), \non \\
 S_2 &=&  -2{a_2\over m_B+m_{\omega}}m_{D'_{1}}f_{D'_{1}}
 A_2^{B\omega}(m_{D'_{1}}^2), \\
 S_3 &=& -2{a_2\over m_B+m_{\omega}} m_{D'_{1}}f_{D'_{1}}
 V^{B\omega}(m_{D'_{1}}^2). \non
 \en
Then the helicity amplitudes $H_0$, $H_+$ and $H_-$ can be
constructed as
 \be
 H_0 &=& {1\over 2m_{D'_1}m_\omega}\left[
 (m_B^2-m_{D'_1}^2-m_\omega^2)S_1+2m_B^2p_c^2S_2\right], \non \\
 H_\pm &=& S_1\pm m_Bp_cS_3.
 \en
The decay rate reads
 \be
 \Gamma(B\to D'_1\omega) &=& {p_c\over 8\pi
 m_B^2}(|H_0|^2+|H_+|^2+|H_-|^2),
 \en
It is found
 \be
 \B(\ov B^0\to D'^0_1\omega)=2.6\times 10^{-4},
 \en
where we have assumed that $B\to\omega$ form factors are the same
as that for $B\to\rho$ transitions which we took from \cite{CCH}.
The above branching ratio prediction is slightly smaller than the
BaBar measurement.

\section{$\ov B\to \ov D_s^{**}D$ decays}

\begin{table}[t]
\caption{Experimental branching ratio products (in units of
$10^{-4}$) of $B$ decays to $D_s^{**}\pi$, where
$D_{s0}^*,D'_{s1}$ stand for the strange charmed mesons
$D_{s0}^*(2317)$ and $D'_{s1}(2460)$, respectively. }
\label{tab:exptBRDsD}
\begin{ruledtabular}
\begin{tabular}{l c c  c}
Mode & BaBar \cite{BaBarDsD} & Belle \cite{BelleDsD} & Average  \\
\hline
 $\B(B^-\to D_{s0}^{*-}D^0)\B(D_{s0}^{*-}\to D_s^{*-}\gamma)$ & &
 $<6.6$ & $<6.6$ \\
 $\B(B^-\to D_{s0}^{*-}D^0)\B(D_{s0}^{*-}\to D_s^{-}\pi^0)$ &
 $10.0\pm3.0\pm1.0^{+4.0}_{-2.0}$ & $9.8^{+2.1}_{-1.9}\pm2.9$ & $9.9^{+2.9}_{-2.6}$ \\
 $\B(B^-\to D_{s0}^{*-}D^{*0})\B(D_{s0}^{*-}\to D_s^{-}\pi^0)$ &
 $9.0\pm6.0\pm2.0^{+3.0}_{-2.0}$ & $<9.3$ & $9.0^{+7.0}_{-6.6}$ \\
 $\B(B^-\to D'^{-}_{s1}D^0)\B(D'^{-}_{s1}\to D_s^-\pi^+\pi^-)$ & &
 $<2.7$ & $<2.7$ \\
 $\B(B^-\to D'^{-}_{s1}D^0)\B(D'^{-}_{s1}\to D_s^-\pi^0)$ & &
 $<2.6$ & $<2.6$ \\
 $\B(B^-\to D'^{-}_{s1}D^0)\B(D'^{-}_{s1}\to D_s^{*-}\pi^0)$ &
 $27\pm7\pm5^{+9}_{-6}$ & $11.6^{+3.9}_{-3.4}\pm3.5$ & $14.1^{+4.7}_{-4.6}$ \\
 $\B(B^-\to D'^{-}_{s1}D^0)\B(D'^{-}_{s1}\to D_s^{*-}\gamma)$ & &
 $<5.8$ & $<5.8$ \\
 $\B(B^-\to D'^{-}_{s1}D^0)\B(D'^{-}_{s1}\to D_s^{-}\gamma)$ & $6.0\pm2.0\pm1.0^{+2.0}_{-1.0}$ &
 $5.9^{+1.1}_{-1.0}\pm1.8$ & $5.9^{+1.7}_{-1.6}$ \\
 $\B(B^-\to D'^{-}_{s1}D^{*0})\B(D'^{-}_{s1}\to D_s^{*-}\pi^0)$ & $76\pm17\pm18^{+26}_{-16}$ &
 $22.3^{+9.8}_{-8.1}\pm6.7$ & $28.0^{+10.7}_{-10.5}$ \\
 $\B(B^-\to D'^{-}_{s1}D^{*0})\B(D'^{-}_{s1}\to D_s^{-}\gamma)$ & $14.0\pm4.0\pm3.0^{+5.0}_{-3.0}$ &
 $9.8^{+3.4}_{-2.9}\pm2.9$ & $11.1^{+3.7}_{-3.5}$ \\
 \hline
 $\B(\ov B^0\to D_{s0}^{*-}D^+)\B(D_{s0}^{*-}\to D_s^{*-}\gamma)$ & &
 $<10.3$ & $<10.3$ \\
 $\B(\ov B^0\to D_{s0}^{*-}D^+)\B(D_{s0}^{*-}\to D_s^{-}\pi^0)$ &
 $18.0\pm4.0\pm3.0^{+6.0}_{-4.0}$ & $10.3^{+2.3}_{-2.0}\pm3.1$ & $12.0^{+3.4}_{-3.3}$ \\
 $\B(\ov B^0\to D_{s0}^{*-}D^{*+})\B(D_{s0}^{*-}\to D_s^{-}\pi^0)$ &
 $15.0\pm4.0\pm2.0^{+5.0}_{-3.0}$ & $5.6^{+2.9}_{-2.3}<13.5$ & $7.1^{+2.7}_{-2.1}$ \\
 $\B(\ov B^0\to D'^{-}_{s1}D^+)\B(D'^{-}_{s1}\to D_s^-\pi^+\pi^-)$ & &
 $<2.7$ & $<2.7$ \\
 $\B(\ov B^0\to D'^{-}_{s1}D^+)\B(D'^{-}_{s1}\to D_s^-\pi^0)$ & &
 $<3.3$ & $<3.3$ \\
 $\B(\ov B^0\to D'^{-}_{s1}D^+)\B(D'^{-}_{s1}\to D_s^{*-}\pi^0)$ &
 $28\pm8\pm5^{+10}_{-~6}$ & $16.9^{+4.3}_{-3.6}\pm5.1$ & $19.3^{+6.0}_{-5.5}$ \\
 $\B(\ov B^0\to D'^{-}_{s1}D^+)\B(D'^{-}_{s1}\to D_s^{*-}\gamma)$ & &
 $<6.3$ & $<6.3$ \\
 $\B(\ov B^0\to D'^{-}_{s1}D^+)\B(D'^{-}_{s1}\to D_s^{-}\gamma)$ & $8.0\pm2.0\pm1.0^{+3.0}_{-2.0}$ &
 $7.1^{+1.4}_{-1.2}\pm2.1$ & $7.4^{+2.1}_{-1.9}$ \\
 $\B(\ov B^0\to D'^{-}_{s1}D^{*+})\B(D'^{-}_{s1}\to D_s^{*-}\pi^0)$ & $55\pm12\pm10^{+19}_{-12}$ &
 $28.8^{+10.1}_{-~8.6}\pm8.6$ & $35.0^{+11.7}_{-10.4}$ \\
 $\B(\ov B^0\to D'^{-}_{s1}D^{*+})\B(D'^{-}_{s1}\to D_s^{-}\gamma)$ & $23\pm3\pm3^{+8}_{-5}$ &
 $14.3^{+3.0}_{-2.7}\pm4.3$ & $17.0^{+4.5}_{-4.0}$ \\
\end{tabular}
\end{ruledtabular}
\end{table}

Under the factorization hypothesis, the decays $\ov B\to \ov
D_s^{**}D$ receive contributions only from the external
$W$-emission diagram, as the penguin contributions to them are
negligible. More precisely, their factorizable amplitudes are
given by
 \be \label{eq:ampBtoDDs}
 A(\ov B\to \ov D_s^{**}D)&=& {G_F\over\sqrt{2}}\Bigg\{\Big(V_{cb}V_{cs}^*\,a_1-V_{tb}V_{ts}^*(a_4+a_{10})\Big)\la
 \ov D_s^{**}|(\bar sc)|0\ra\la D|(\bar c b)|\ov B\ra \non \\
 && +2V_{tb}V^*_{ts}(a_6+a_8)\la \ov D_s^{**}|\bar
 s(1+\gamma_5)c|0\ra\la D|\bar c(1-\gamma_5)b|\ov B\ra\Bigg\},
 \en
where $(\bar q_1q_2)\equiv \bar q_1\gamma_\mu(1-\gamma_5)q_2$.
Since the tensor meson cannot be produced from the $V-A$ current,
the $B$ decay into $\ov D_{s2}^*D^{(*)}$ is prohibited under the
factorization hypothesis. Except for $D_{s2}^*$, the measurement
of $\ov B\to \ov D_s^{**}D$ can be used to determine the decay
constant of $D_s^{**}$ to the approximation that penguin
contributions are neglected.

The current measurements of $\ov B\to \ov D_s^{**}D$  are
summarized in Table \ref{tab:exptBRDsD}. They are consistent with
relation $\Gamma(\ov B^0\to D_s^{**-}D^+)=\Gamma(B^-\to
D_s^{**0}D^+)$, but it is obvious that one needs more improved
data to test the above relation.

Since $D_{s0}^*(2317)$ is below the $DK$ threshold, the only
allowed hadronic decay is the isospin-violating one, namely,
$D_s^+\pi^0$. Thus far, no other modes have been observed (see
e.g. \cite{BaBarDs0Ds1}). The allowed radiative mode is
$D_{s0}^*\to D_s^*\gamma$, which is constrained to be
 $\Gamma(D_{s0}^*\to D_s^*\gamma)/\Gamma(D_{s0}^*\to
D_s\pi^0)<0.059$ by CLEO \cite{CLEODs1}. Therefore,
$0.94\lsim\B[D_{s0}^*(2317)\to D_s\pi^0]\lsim 1.0$. It follows
from Table \ref{tab:exptBRDsD} and ({\ref{eq:ampBtoDDs}) that
 \be \label{eq:fDs0}
 a_1f_{D_{s0}^*}= \cases{58\sim83~{\rm MeV} & from~$B^-$~decays, \cr
 63\sim86~{\rm MeV} & from~$\ov B^0$~decays. }
 \en
Hence, our prediction $f_{D_{s0}^*}=71$ MeV is in good agreement
with experiment.

\begin{table}[t]
\caption{The predicted branching ratios for $D'^+_{s1}(2460)$.
Branching ratios in parentheses are with respect to the decay mode
$D_s^{*+}\pi^0$, so are the experimental results.}
\label{tab:BRDs2460}
\begin{ruledtabular}
\begin{tabular}{l l l  l l  r r r }
Mode & \cite{Bardeen} & \cite{Godfrey} & BaBar
\cite{BaBarDs0Ds1,BaBarDsD}~\footnotemark[1]
 & Belle \cite{BelleDs1,BelleDsD} & CLEO \cite{CLEODs1}  \\
\hline
 $D_s^{*+}\pi^0$ & 56\% & 43\% &  \\
 $D_s^+\gamma$ & 13\%(0.24) & 27\%(0.63) &  $0.337\pm0.036\pm0.038$~\footnotemark[2] & $0.63\pm0.15\pm0.15$
  & $<0.49$\\
 & & & $0.275\pm0.045\pm0.020$~\footnotemark[3] & $0.43\pm0.08\pm0.04$ \\
 $D_s^{*+}\gamma$ & 12\%(0.22) & 24\%(0.56) &  $<0.24$ & $<0.31$ & $<0.16$ \\
 $D^+_s\pi^+\pi^-$ & 11\%(0.20) & 7\%(0.16) &   $0.077\pm0.013\pm0.008$ & $<0.13$ & $<0.08$ \\
 $D_{s0}^{*+}\gamma$ & 7\%~\,(0.13) & 0.05\%($1.2\times 10^{-3})$ &  $<0.25$ & & $<0.58$ \\
 $D_{s0}^{*+}\pi$ & 0.002\% & &
\end{tabular}
\end{ruledtabular}
\footnotetext[1]{The BaBar results are for the branching ratios
with respect to $D'_{s1}\to D_s\pi^0\gamma$ arising from
$D_s^*\pi^0$ and $D^*_{s0}\gamma$. The BaBar data are consistent
with the decay $D'_{s1}\to D_s\pi^0\gamma$ proceeding entirely
through $D_s^*\pi^0$.}
 \footnotetext[2]{from continuum $e^+e^-\to c\bar
c$.}
 \footnotetext[3]{from $B$ decay.}
\end{table}

Various model predictions on the branching ratios of
$D'_{s1}(2460)$ are shown in Table \ref{tab:BRDs2460}. In general,
the models \cite{Godfrey} and \cite{Bardeen} differ in the
predictions of the branching fractions for the hadronic and
radiative decays of $D'_{s1}$ which in turn will lead to different
absolute branching ratios for $\ov B\to D'_{s1}D$. Very recently
BaBar has been able to measure $\B(\ov B\to D'_{s1}D)$ without any
assumption on the decays of $D'_{s1}$ \cite{BaBarBDsDBR}. What
BaBar have measured are the decays $\ov B\to D_{\rm meas}D_X$ with
$D_{\rm meas}$ being a fully reconstructed charmed meson. The mass
and momentum of the $D_X$ are then inferred from the kinematics of
the two-body $B$ decay. By selecting final states with
$D_X=D_{s1}(2460)^-$, BaBar measurements of $\B(\ov B\to
D'_{s1}D^{(*)})$ are summarized in Table \ref{tab:BaBarBDsD}.
Combining with the BaBar results for branching ratio products of
$B$ decays to $D'_{s1}\pi$ (Table \ref{tab:exptBRDsD}),  the
absolute branching ratios of $D'_{s1}(2460)$ are determined to be
\cite{BaBarBDsDBR}
 \be \label{eq:exptBRDs1}
 \B(D'_{s1}\to D_s^*\pi^0)=0.56\pm0.13\pm0.09,\qquad
 \B(D'_{s1}\to D_s\gamma)=0.16\pm0.04\pm0.03\,.
 \en
It is clear from Table \ref{tab:BRDs2460} that model
\cite{Bardeen} is more preferred. It is interesting to note that
the sum $\B(D'_{s1}\to D_s^*\pi^0+D_s\gamma)\approx 0.70$ turns
out to be the same in both models and is consistent with the
measurement $0.72\pm0.17$ derived from Eq. (\ref{eq:exptBRDs1}).

\begin{table}[t]
\caption{Branching ratios (in units of $10^{-4}$) of $B$ decays to
$D'_{s1}D^{(*)}$ measured by BaBar \cite{BaBarBDsDBR}.}
\label{tab:BaBarBDsD}
\begin{center}
\begin{tabular}{|l c| c c|}
\hline
~~~~~Mode~~~~~ & ~~~~~~Br~~~~~~~~ & ~~~~~~~~Mode~~~~~~~~ & ~~~~~~~~Br~~~~~~~~   \\
\hline
 $B^-\to D'^-_{s1}D^{0}$ & $43\pm16\pm13$ & $B^-\to
 D'^-_{s1}D^{*0}$ & $112\pm26\pm20$ \\
 $\ov B^0\to D'^-_{s1}D^{+}$ & $26\pm15\pm7$ & $\ov B^0\to
 D'^-_{s1}D^{*+}$ & $88\pm20\pm14$ \\
 \hline
\end{tabular}
\end{center}
\end{table}

 The decay constant of $D'_{s1}$ is then found to be
 \be \label{eq:fDs1}
 a_1f_{D'_{s1}}=\cases{188^{+40}_{-54}~{\rm MeV} & from~$B^-$~decays, \cr
 152^{+43}_{-62}~{\rm MeV} & from~$\ov B^0$~decays. }
 \en
For comparison, the decay constants
 \be
 a_1f_{D^*_{s0}}> 74\pm 11~{\rm MeV}, \qquad a_1f_{D'_{s1}}>166\pm20~{\rm MeV}
 \en
are obtained in \cite{Hwang}. Comparing (\ref{eq:fDs1}) with
(\ref{eq:fDs0}) we see that $f_{D'_{s1}}$ is about two times of
$f_{D^*_{s0}}$, whereas $f_{D^*_0}$ and $f_{D'_1}$ are very
similar in size [cf Eqs. (\ref{eq:fDs1}) and (\ref{eq:fD1})]. It
is not clear to us why the HQS relation for the decay constants is
satisfied for the sector $\{D_0^*,D'_1\}$ but badly broken for
$\{D_{s0}^*,D'_{s1}\}$.

The decays $\ov B\to D_s^{**-}D$ have also been studied in
\cite{Thomas}. However, the branching ratios $\B(\ov B\to
D_{s0}^{*-}D)=(3.0\sim 3.8)\times 10^{-3}$, $\B(\ov B\to
D'^{-}_{s1}D)=(7.2\sim 8.9)\times 10^{-3}$ and $\B(\ov B\to
D'^{-}_{s1}D^*)=(2.5\sim 2.9)\times 10^{-2}$ predicted in
\cite{Thomas} are too large compared to experiment (see Tables
\ref{tab:exptBRDsD} and \ref{tab:BaBarBDsD}).

For decays $B^-\to D_s^{**-}D^{*0}$, their factorizable amplitudes
are given by
 \be
 A(B^-\to D_{s0}^{*-}D^{*0}) &=& 2(\vp^*\cdot p_B)a_1
 f_{D_s^*}m_{D^*}A_0^{BD^*}(m_{D_{s0}^*}^2),   \\
 A[B^-\to D'^{-}_{s1}(\vp_{D_{s1}},p_{D_{s1}})D^{*0}(\vp_{D^*},p_{D^*})]
 &=&
\vp_{D_1}^{*\mu}\vp_{D^*}^{*\nu}[S_1 g_{\mu\nu}+S_2(p_B)_\mu
(p_B)_\nu+iS_3\epsilon_{\mu\nu\alpha\beta}p_{D_{s1}}^\alpha
p_{D^*}^\beta], \non
 \en
with
 \be \label{S123}
 S_1 &=& a_1(m_B+m_{D^*})m_{D'_{s1}}f_{D'_{s1}}
 A_1^{BD^*}(m_{D'_{s1}}^2), \non \\
 S_2 &=&  -2{a_1\over m_B+m_{D^*}}m_{D'_{s1}}f_{D'_{s1}}
 A_2^{BD^*}(m_{D'_{s1}}^2), \non \\
 S_3 &=& -2{a_1\over m_B+m_{D^*}} m_{D'_{s1}}f_{D'_{s1}}
 V^{BD^*}(m_{D'_{s1}}^2).
 \en
We consider the ratios $\Gamma(\ov B\to D_s^{**-}D^*)/\Gamma(\ov
B\to D_s^{**-}D)$ for $D_s^{**}=D_{s0}^*(2317)$ and
$D'_{s1}(2460)$ which have the advantage of being
$a_1f_{D_s^{**}}$ independent. The results are exhibited in Table
\ref{tab:ratioDsD} where uses of the $B\to D^*$ form factors from
\cite{CCH} have been made. It is evident that the predictions are
consistent with experiment.

\begin{table}[t]
\caption{The predicted ratios $\Gamma(\ov B\to
D_s^{**}D^*)/\Gamma(\ov B\to D_s^{**}D)$ for $D_s^{**}=D_{s0}^*$
and $D'_{s1}$. Experimental results are taken from Tables
\ref{tab:exptBRDsD} and \ref{tab:BaBarBDsD}.}
 \label{tab:ratioDsD}
\begin{center}
\begin{tabular}{|c c c|} \hline
Mode & ~~~~~Theory~~~~~ & ~~~~~Expt~~~~~   \\
\hline
 $D_{s0}^{*-}D^{*0}/D_{s0}^{*-}D^0$ & 0.49 & $0.91\pm0.73$  \\
 $D_{s0}^{*-}D^{*+}/D_{s0}^{*-}D^+$ & 0.49 & $0.59\pm0.26$  \\
 $D'^-_{s1}D^{*0}/D'^-_{s1}D^0$ & 3.6 & $3.4\pm2.4$  \\
 $D'^-_{s1}D^{*+}/D'^-_{s1}D^+$ & 3.6 & $2.6\pm1.5$  \\
 \hline
\end{tabular}
\end{center}
\end{table}

\section{$\ov B\to D_s^{**}K$ decays}

\begin{table}[t]
\caption{Experimental branching ratio products (in units of
$10^{-4}$) of $B$ decays to $D^{**+}_sK^-$ and $D_s^{**-}\pi^+$
where $D_{s0}^*,D'_{s1}$ stand for the strange charmed mesons
$D_{s0}^*(2317)$ and $D'_{s1}(2460)$, respectively. }
\label{tab:exptBRDsK}
\begin{ruledtabular}
\begin{tabular}{l c c  c}
Mode & BaBar \cite{BaBarDspi} & Belle \cite{BelleDsK} & Average  \\
\hline
 $\B(\ov B^0\to D_s^+K^-)$ & 0$.25\pm0.04\pm0.04$ &
 $0.24^{+0.10}_{-0.08}\pm0.7$ & $0.25\pm0.06$ \\
 $\B(\ov B^0\to D_s^-\pi^+)$ & $0.13\pm0.03\pm0.02$ &
 $0.46^{+0.12}_{-0.11}\pm0.13$ & $0.14\pm0.04$ \\
 \hline
 $\B(\ov B^0\to D_{s0}^{*+}K^-)\B(D_{s0}^{*+}\to D_s^+\pi^0)$ & &
 $0.44\pm0.08\pm0.06\pm0.11$ & $0.44\pm0.15$ \\
 $\B(\ov B^0\to D'^+_{s1}K^-)\B(D'^+_{s1}\to D_s^+\gamma)$ & &
 $<0.086$ & $<0.086$ \\
 \hline
 $\B(\ov B^0\to D_{s0}^{*-}\pi^+)\B(D_{s0}^{*-}\to D_s^-\pi^0)$ & &
 $<0.25$ & $<0.25$ \\
 $\B(\ov B^0\to D'^-_{s1}\pi^+)\B(D'^-_{s1}\to D_s^-\gamma)$ & &
 $<0.04$ & $<0.04$ \\
\end{tabular}
\end{ruledtabular}
\end{table}

The experimental branching ratio products of $\ov B^0$ decays to
$D^{**+}_sK^-$ and $D_s^{**-}\pi^+$ are summarized in Table
\ref{tab:exptBRDsK}. The decay $\ov B^0\to D^{**+}_sK^-$ can only
proceed through the $W$-exchange diagram, whereas $\ov B^0\to
D^{**-}_s\pi^+$ does receive an external $W$-emission contribution
which is however CKM suppressed. From Table \ref{tab:exptBRDsK} it
is obvious that $\Gamma(\ov B^0\to D_{s0}^{*+}K^-)\gsim \Gamma(\ov
B^0\to D_s^+K^-)$. These two decays can only proceed via a
short-distance $W$-exchange process or through the long-distance
final-state rescattering processes $\ov B^0\to D^+\pi^-\to D_s^+
K^-$ and $\ov B^0\to D^{*+}_0\pi^-\to D^{*+}_{s0} K^-$. (In fact,
the rescattering process has the same topology as $W$-exchange.)
Since $\B(\ov B^0\to D^+\pi^-)\approx 2.8\times 10^{-3}\gg \B(\ov
B^0\to D_s^+K^-)$, it is thus expected that the decay $\ov B^0\to
D_s^+ K^-$  is dominated by the long-distance rescattering
process. As $\B(\ov B^0\to D_0^{*+}\pi^-)<1.8\times 10^{-4}$
\cite{PDG}, we will naively conclude that $\Gamma(\ov B^0\to
D_{s0}^{*+}K^-)/\Gamma(\ov B^0\to D_s^+K^-)\approx \Gamma(\ov
B^0\to D_0^{*+}\pi^-)/\Gamma(\ov B^0\to D^+\pi^-) < 0.06$, where
we have assumed that the rescattering effects of $D^+\pi^-\to
D_s^+ K^-$ and $D^{*+}_0\pi^-\to D^{*+}_{s0} K^-$ are similar.
This is obviously in contradiction to experiment. Nevertheless, if
$D_{s0}^*(2317)^+$ is a bound state of $c\bar sd\bar d$~\cite{CH},
then a tree diagram will contribute and this may allow us to
understand why $\Gamma(\ov B^0\to D_{s0}^{*+}K^-)\gsim \Gamma(\ov
B^0\to D_s^+K^-)$.

For $\ov B^0\to D_s^{**-}\pi^+$ decays, we obtain
 \be
 \B(\ov B^0\to D_{s0}^{*-}\pi^+)=(2.6^{+0.9}_{-0.7})\times 10^{-6}, \qquad\quad
 \B(\ov B^0\to D'^{-}_{s1}\pi^+)=(1.5\pm0.4)\times 10^{-5}.
 \en
They are consistent with the experimental limits: $\B(\ov B^0\to
D_{s0}^{*-}\pi^+)<2.5\times 10^{-5}$ and $\B(\ov B^0\to
D'^{-}_{s1}\pi^+)<2.7\times 10^{-5}$, where use has been made of
Eq. (\ref{eq:exptBRDs1}).

\section{conclusions}

We have studied the production of even-parity charmed mesons in
hadronic $B$ decays, namely, the Cabibbo-allowed decays $\ov B\to
D^{**}\pi$ and $\bar D^{**}_sD^{(*)}$. The main conclusions are
the following:

\begin{itemize}

\item We have shown that the measured color-allowed decays $\ov
B^0\to D^{**+}\pi^-$ are consistent with the theoretical
expectation. However, the experimental observation of $B^-\to
D^{**0}\pi^-$ for the broad $D^{**}$ states is rather astonishing
as it requires a color-suppressed decay amplitude be larger than
the color-allowed one, although the former is  $1/m_b$ suppressed
in the heavy quark limit.

\item In order to accommodate the data of $\ov B\to D^{**}\pi$, it
is found that the real part of $a_2/a_1$ with a sign opposite to
that in $\ov B\to D\pi$ decays, where $a_1$ and $a_2$ are the
effective parameters for color-allowed and color-suppressed decay
amplitudes, respectively. This indicates that $a_2$ is process
dependent and the nonfactorizable contributions to
color-suppressed amplitudes are not universal; that is, they are
process dependent.

\item The decay constants and form factors for $D^{**}$ and the
Isgur-Wise functions $\tau_{1/2}(\omega)$ and $\tau_{3/2}(\omega)$
are extracted from the $B\to D^{**}\pi$ data. The Isgur-Wise
functions calculated in the covariant light-front quark model are
in good agreement with experiment.

\item The color-suppressed modes $\ov B^0\to D^{**0}\pi^0$ for
broad $D^{**}$ states and $\ov B^0\to D'^0_1(2430)\omega$ are
predicted to have branching ratios of order $10^{-4}$. Robust
lower bounds on $\ov B^0\to\{ D_0^*,D'^0_1\}\pi^0$ can be derived
from the isospin argument.

\item The decay constants of $D_{s0}^*(2317)$ and $D'_{s1}(2460)$
are inferred from the measurements of $\ov B\to D_s^{**-}D$ to be
$58\sim 86$ MeV and $130\sim 200$ MeV, respectively. The large
disparity between $f_{D_{s0}}$ and $f_{D'_{s1}}$ is surprising and
unexpected. The ratios $\Gamma(\ov B\to D_{s0}^{*-}D^*)/\Gamma(\ov
B\to D_{s0}^{*-}D)$ smaller than unity and $\Gamma(\ov B\to
D_{s1}^{-}D^*)/\Gamma(\ov B\to D_{s1}^{-}D)$ larger than unity are
confirmed by experiment.

\item The observation that the production of $D_{s0}^{*+}K^-$ is
larger than $D_s^+K^-$ may imply a four-quark structure for the
scalar charmed meson $D_{s0}^*$.

 \end{itemize}

There are two main puzzles concerning $\ov B\to D^{**}\pi$ decays,
namely, why the color-suppressed amplitude dominates in the broad
$D^{**}$ production in charged $B$ decays and why is the sign of
$a_2/a_1$ different from that in $\ov B\to D\pi$ decays ? Thus
far, we have analyzed the $\ov B\to D^{**}\pi$ and $D_s^{**-}D$
decays within the phenomenological framework of generalized
factorization. In the QCD factorization approach, the parameter
$a_2$ is not calculable owing to the presence of infrared
divergence caused by the gluon exchange between the emitted
$D^{**}$ meson and the $(\ov B\pi)$ system. That is, the
nonfactorizable contribution to $a_2$ is dominated by
nonperturbative effects.

In soft-collinear effective theory (SCET), the color-suppressed
$\ov B^0\to D^{**0}\pi^0$ decay is proved to be factorizable. More
precisely, its amplitude factors into a pion light-cone wave
function and a $B\to D^{**}$ soft distribution function rather
than being like the naive $a_2$ factorization as shown in Eq.
(\ref{eq:BtoD0pi0}) \cite{Mantryb,Mantry}. However, it is
difficult to make a quantitative prediction in SCET at this stage
and it is not clear to us if SCET can provide an explanation of
why the sign of $a_2/a_1$ flips from $\ov B\to D\pi$ to $\ov B\to
D^{**}\pi$.

\vskip 2.0cm \acknowledgments
 We are grateful to Iain Stewart for discussions.
This research was supported in part by the National Science
Council of R.O.C. under Grant Nos. NSC94-2112-M-001-016,
NSC94-2112-M-001-023 and NSC94-2811-M-001-059.


\end{document}